\documentclass[aps,prd,preprint,superscriptaddress,showpacs,showkeys]{revtex4}
\usepackage{amssymb,amsmath}
\usepackage{graphicx} 
\usepackage{dcolumn} 

\usepackage{epsfig}   
\usepackage{pstricks}
\usepackage{ifpdf}

\newcommand{\tr}{ {\mathrm{tr}\, }}
\newcommand{\Tr}{ {\mathrm{Tr}\, }}

\begin{document}

\title{ QCD Effective Locality : A Theoretical and Phenomenological Review }

\author{H.M. Fried}
\affiliation{Physics Department, Brown University, Providence, RI 02912, USA}
\email[]{herbert_fried@brown.edu}
\author{Y. Gabellini}
\affiliation{Universit\'{e} C\^ote d'Azur\\ Institut de Physique de Nice, UMR CNRS 7010 1361 routes des Lucioles, 06560 Valbonne, France}
\email[]{Yves.Gabellini@inphyni.cnrs.fr}
\author{T. Grandou}
\affiliation{Universit\'{e} C\^ote d'Azur\\ Institut de Physique de Nic{e}, UMR CNRS 7010 1361 routes des Lucioles, 06560 Valbonne, France}
\email[]{Thierry.Grandou@inphyni.cnrs.fr}
\author{P.H. Tsang}
\affiliation{Physics Department, Brown University, Providence, RI 02912, US{A}}
\email[]{Peter_Tsang@brown.edu}

\date{\today}
\vspace{5cm}

\begin{abstract} 
About ten years ago the use of standard functional manipulations was demonstrated to imply an unexpected property satisfied by the fermionic Green's functions of QCD and dubbed \textit{Effective Locality}. This feature of QCD is non--perturbative as it results from a full \emph{gauge invariant} integration of the gluonic degrees of freedom. 
In this review article, a few salient theoretical 	aspects and phenomenological applications of this property are summarized. 
\end{abstract}


\maketitle


 \section{\label{SEC:1}Introduction}
 Over the past decade a number of articles has been devoted to the study of a new property concerning the non--perturbative regime of $QCD$ \cite{QCD1,QCD-II,QCD5, QCD6, QCD5'}.  This property, of the non--perturbative fermionic Green's functions of $QCD$, has been named {\textit{Effective Locality}}, or $EL$ for short. In words, the $EL$ property can be phrased as follows. 
 \par
\emph{For any fermionic $2n$--point Green's functions, the
full gauge--fixed sum of cubic and quartic gluonic interactions,
fermionic loops included, results in a local contact--type interaction. This local interaction is
mediated by a tensorial field which is antisymmetric both in Lorentz and
colour indices. Moreover, the resulting sum appears to be fully gauge--fixing independent.}
\par
Because the integrations of elementary degrees of freedom usually result in highly non--local 
 effective interactions, this resulting Effective Locality is rather surprising and is suggestive of a form of {\textit{duality}}.\par
 In the pure euclidean Yang--Mills case in effect, and at first non trivial order of a semi--classical expansion, this same phenomenon of a resulting local effective interaction has been noticed in an attempt to find a formulation \emph{dual} to the original Yang--Mills theory \cite{RefF}.  The case of $QCD$ is different, though. Apart from a \textit{supersymmetric} extension \cite{SW1994}, $QCD$ is not known to admit any dual formulation and  Effective Locality calculations themselves also attest to this situation. It remains that Effective Locality calculations proceed from first principles and offer a useful means to learn about non--perturbative physics in $QCD$. 

\par\medskip\medskip
The current paper aims at offering a concise review of the $EL$ outputs both at theoretical and phenomenological levels. In the next Section the basics of the $EL$ property will accordingly be recalled, while Section III will summarise its theoretical aspects. The phenomenological calculations which have been achieved out the $EL$ property are presented in Section IV and a concise conclusion about these ongoing analyses is given in Section V.
\section{\label{SEC:2}Effective locality in short}
\subsection{The 4--point fermionic Green's function}
Effective Locality  is derived in the context of $QCD$, with the help of standard functional techniques. Thus, one starts with the $QCD$ lagrangian, to which, is added and subtracted one and the same covariant density of $(1/ 2\zeta)\,(\partial^{\mu}A_{\mu}^{a})^2$,
\begin{equation}\label{1}
\mathcal{L}_{\mathrm{QCD}} = \bar{\psi}\,( i\,\slash \!\!\!\partial - m)\,\psi  - \frac{1}{4} {F}_{\mu \nu}^{a} {{F}^{\mu \nu}}^{a} - \,{1\over 2\zeta}\,(\partial^{\mu}A_{\mu}^{a})^2+{1\over 2\zeta}\,(\partial^{\mu}A_{\mu}^{a})^2+ g \,\bar{\psi}\,\slash \!\!\!\! A^{a} \displaystyle{\lambda^{a}\over2}\, \psi 
\end{equation}
In order to keep things as simple as possible, one flavour of quark is considered only. The eight $\lambda^as$ are the $SU_c(3)$ Lie algebra generators taken in the fundamental representation~: $\displaystyle [\,{\lambda^a\over2},{\lambda^b\over2}\,] = if^{abc}\,{\lambda^c\over2}$, where the totally antisymmetric structure constants $f^{abc}$ will play a key role in what follows. The gluon field tensor is $\displaystyle{F}_{\mu\nu}^a\equiv \partial_{\mu}A_{\nu}^a - \partial_{\nu}A_{\mu}^a + gf^{abc}A_{\mu}^bA_{\nu}^c$. Concerning the apparent `gauge fixing term' $\displaystyle{1\over 2\zeta}\,(\partial_{\mu}A^{\mu})^2$, one will choose the Feynman gauge at $\zeta = 1$ for the sake of illustration.

The metric used in this article is the usual $g_{\mu\nu} = (\,+, -, -, -\,)$. There can result a few sign discrepancies, of no consequences, with the original papers \cite{QCD1,QCD-II,QCD5, QCD6, QCD5'}, where the so-called West Coast metric was employed.

In the articles \cite{QCD1,QCD-II,QCD5, QCD6, QCD5'}, the $EL$ property shows up at the level of fermionic Green's functions, given by functional differentiations of the subsequent $QCD$ generating functional ${Z}_{\mathrm{QCD}}[\,j, \bar{\eta}, \eta\,]$, taken with respect to the quark sources $\bar{\eta}$, $\eta$, thereafter sent to zero. For configuration--space amplitudes, one has,
\begin{equation}\label{amp} \mathbf{M}(x_{1}, y_{1}; x_{2}, y_{2}) = \left. \frac{\delta}{\delta \bar{\eta}(x_{1})}\frac{\delta}{\delta \bar{\eta}(x_{2})}\frac{\delta}{\delta \eta(y_{1})}\frac{\delta}{\delta \eta(y_{2})}{Z}_{\mathrm{QCD}}[\, j, \bar{\eta}, \eta\,] \right|_{\displaystyle\eta=\bar\eta=0; j=0}\end{equation}
where (see \cite{QCD1}) :
\begin{eqnarray}\label{Z}
&{Z}_{\mathrm{QCD}}[\,j, \bar{\eta}, \eta\,] = \mathcal{N}\,e\,^{\displaystyle-{ i\over2}\int\!\!d^4x\,d^4y\,j^a_{\mu}(x)\,D_{\mathrm{F}\mu\nu}^{(0)ab}(x-y)\, j^b_{\nu}(y)}\, \nonumber \\&\, \times\,e\,^{\displaystyle{i\over2}\int\!\!d^4x\,d^4y\,{\delta\over\delta A_{\mu}^a(x)}\,D_{\mathrm{F}\mu\nu}^{(0)ab}(x-y)\,{\delta\over\delta A_{\nu}^b(y)} }\\ \nonumber&\times\,   e^{\displaystyle-\frac{i}{4} \int\!\! dx\,{F}_{\mu \nu}^{a}(x) {{F}^{\mu \nu}}^{a}(x) - \frac{i}{2} \int\!\! dx\,{ A_{\mu}^a(x) \,\partial^{2}A_{\mu}^a(x)} } \,e^{\displaystyle -i\!\int\!\!d^4x\,d^4y\,\bar\eta(x)\, G_F(x,y|A)\,\eta(y)}\, e\,^{\displaystyle{L}[A] }\hfill\hfill
\end{eqnarray}
with $\displaystyle A_{\mu}^a(x) = \int\!\!d^4y\,D_{\mathrm{F}\mu\nu}^{(0)ab}(x-y)\,j^{b}_{\nu}(y)$, and where  $D_{\mathrm{F}\mu\nu}^{(0)ab}$ is the free gauge field propagator in the Feynman gauge, $\partial^2D_{\mathrm{F}\mu\nu}^{(0)ab}= \delta^{ab}\,g_{\mu\nu}1\kern -2.5pt{\rm l}$. The normalisation factor $\mathcal{N}$ is such that ${Z}_{\mathrm{QCD}}[\,0, 0, 0\,] = 1$, and for the quark propagator $G_F(x,y|A)$, a \emph{Fradkin representation} can be used to make explicit the non--abelian gauge--field dependences (\cite{Herb} chap.3) :
\begin{eqnarray} \label{Fradkin}& \displaystyle G_F(x,y|A) =- i\,{\cal N}\int_0^{\infty}\!\!ds\,e\,^{\displaystyle -ism^2}\int\!d[u_{\mu}]\,e\,^{\displaystyle-{i\over4}\int\!\!\!\int_0^s\!\!ds_1ds_2\, u_{\mu}(s_1)\,h^{-1}(s_1, s_2)\, u_{\mu}(s_2)}\nonumber\\ \noalign{\smallskip} & \times\displaystyle \,\bigl(\, i\gamma_{\mu}\,{\delta\over\delta u_{\mu}'(s)} + m \bigr)\,\delta^{(4)}(x - y + u(s))\\ \noalign{\smallskip} & \times\,T_{s'}\,e\,^{\displaystyle -ig\!\int_0^s\!\!ds'\, u_{\mu}'(s')\,A_{\mu}^a(y-u(s')){\lambda^{a}\over2}+ ig\!\int_0^s\!\!ds'\,\sigma_{\mu\nu}\,F_{\mu\nu}^a(y-u(s')){\lambda^{a}\over2}}\nonumber\hfill
\end{eqnarray}

In (\ref{Fradkin}), $u_\mu(s)$ stands for the Fradkin field variable, $T_{s'}$ indicates a prescription of $s'$ \emph{proper--time} ordering \cite{IZ}, the $\sigma_{\mu\nu}=\displaystyle{i\over4}\,[\gamma_\mu,\gamma_\nu]$ are the usual Lorentz group generators for spinors. 

One has also,  
$\displaystyle h^{-1}(s_1, s_2) = {\overrightarrow\partial\over\partial  s_1}\delta(s_1-s_2){\overleftarrow\partial\over\partial  s_2}$, so that,
$$\int\!\!\!\int_0^s\!\!ds_1ds_2\, u_{\mu}(s_1)\,h^{-1}(s_1, s_2)\, u_{\mu}(s_2) = \int_0^s\!\!ds'\, (u{\,}'(s'))^ 2$$ 
The constant ${\cal N}$ is such that $\displaystyle{\cal N}^{-1} =  \displaystyle\int\!d[u]\,e\,^{\displaystyle-{i\over4}\int\!u\, h^{-1}\,u} = \int\!d[u]\,e\,^{\displaystyle-{i\over4}\int\!u\,'\,^2}$ 

The fermionic closed loop functional $L[A] = \displaystyle{\rm Tr}\ln(\,1\hskip -3truept {\rm I}-e\,\,\slash \!\!\!\!A^a\,{\lambda^{a}\over2}S_F)$ appearing in (\ref{Z}), with $S_F$ the free fermion propagator, can also be endowed with a Fradkin representation similar to the propagator $G_F[A]$, see \cite{Herb} chap.3 and eq.(\ref{Eq:LFradkin01}) below.

To be able to compute $\mathbf{M}$ and carry out $EL$ calculations one introduces for the $\displaystyle\int\!\! F^2$ term which appears in the right hand sides of (\ref{Z}) an auxiliary $\chi$--field \cite{RefF,Halpern1977a,Halpern1977b} :
\begin{equation}\label{chi}
\displaystyle e\,^{\displaystyle - {i\over 4}\int\!\!d^4x\, {F}_{\mu\nu}^a(x)\,{F}^{a\mu\nu}(x)} = \displaystyle\mathcal{N}' \int\!\!d[\chi]\,e\,^{\displaystyle  {i\over 4}\int\!\!d^4x\, ({\chi}_{\mu\nu}^a(x))^2  + {i\over 2}\int\!\!d^4x\, {\chi}_{\mu\nu}^a(x)\,{F}^{a\mu\nu}(x)}
\end{equation}
where ${\chi}_{\mu\nu}^a(x)$ is antisymmetric in $\mu$ and $\nu$, like ${F}_{\mu\nu}^a(x)$, and where the measure is,

\begin{equation}\int{\mathrm{d}[\chi]} = \prod_{z} \prod_{a} \prod_{\mu <\nu} \int{\mathrm{d}[\chi_{\mu \nu}^{a}(z)}]
\end{equation}
Spacetime is taken as a collection of cells of infinitesimal size $\delta^{4}$ about each point $z$ \cite{RefO}, and $\mathcal{N}'$ is a normalisation constant $ \displaystyle\mathcal{N'} \!\!\int\mathrm{d}[\chi] \, e\,^{\displaystyle{ \frac{i}{4} \int\!\chi^{2}}} \!\!= 1$. The generating functional (\ref{Z}) can therefore be rewritten as ($\mathcal{N}' \cdot \mathcal{N} = \mathcal{N}'' \equiv \mathcal{N}$) :
\begin{eqnarray}\label{Z1}
&{}&{Z}_{\mathrm{QCD}}[j,\bar{\eta},\eta] = \mathcal{N}\,e\,^{\displaystyle-{ i\over2}\int\!\!d^4x\,d^4y\,j^a_{\mu}(x)\,D_{\mathrm{F}\mu\nu}^{(0)ab}(x-y)\, j^b_{\nu}(y)}\nonumber\\ &\times &\int\!\mathrm{d}[\chi] \, e\,^{\displaystyle  {i\over 4}\int\!\!d^4x\, ({\chi}_{\mu\nu}^a(x))^2} \,  e\,^{\displaystyle{\mathfrak{D}_{A}^{(0)}}}\, e\,^{\displaystyle  {i\over 2}\int\!\!d^4x\, {\chi}_{\mu\nu}^a(x)\,{F}^{a\mu\nu}(x)} \\ &  \nonumber \times &e^{\displaystyle - \frac{i}{2} \int\!\! dx\,{ A_{\mu}^a(x) \,\partial^{2}A_{\mu}^a(x)} } \,e^{\displaystyle -i\!\int\!\!d^4x\,d^4y\,\bar\eta(x)\, G_F(x,y|A)\,\eta(y)}\, e\,^{\displaystyle{L}[A] }\biggl|_{\displaystyle A = \int\!\!{{D}_{\mathrm{F}}^{(0)}j} }
\end{eqnarray}
\noindent by using the shorthand notation,  
\begin{equation}\label{link}\mathfrak{D}_{A}^{(0)} = {\displaystyle{i\over2}\,\int\!\!d^4x\,d^4y\,{\delta\over\delta A_{\mu}^a(x)}\,D_{\mathrm{F}\mu\nu}^{(0)ab}(x-y)\,{\delta\over\delta A_{\nu}^b(y)} }\end{equation} 
and where the functional integration on $\chi$ and the functional differentiations operated by the so--called \emph{linkage operator}, $\exp\,\mathfrak{D}_{A}^{(0)}$, can be permuted without prejudice.

\par\smallskip

\par
\par

One will deal with a 4--point fermionic Green's function (\ref{amp}) as, through more cumbersome expressions, the following structures easily extend to the $2n$--point general case (see \cite{RefI}, App.~D). Then, two propagators $G_F^{1}(x_1,y_1|A)$ and $G_F^{2}(x_2,y_2|A)$ appear, represented by (\ref{Fradkin})~:
\begin{eqnarray}\label{4pts}
& & \mathbf{M}(x_{1}, y_{1}; x_{2}, y_{2}) = \mathcal{N} \, \int\!\!{d[\chi] \, e\,^{\displaystyle{\frac{i}{4}\int{\chi^{2}}}} \, e\,^{\displaystyle{\mathfrak{D}_{A}^{(0)}}} \,}\, e\,^{\displaystyle{\frac{i}{2}\int{\chi\,F} -\frac{i}{2}\int{A\, \partial^2 A }} }\\ \nonumber &\times &  G_F^1(x_{1}, y_{1}|A) \, G_F^2(x_{2}, y_{2}|A)  \, e\,^{\displaystyle{L[A]}}\Bigl|_{\displaystyle{A=0}}
\end{eqnarray}
\indent In (\ref{Z1}), note that a direct replacement of $G_F[A]$ by the expression (\ref{Fradkin}) would result in an involved structure of an exponential of an exponential. Functional differentiations with respect to the \emph{Grassmannian sources} $\bar{\eta},\eta$ allow to circumvent this complication and it is therefore on fermionic Green's functions that the $EL$ property has been first disclosed. Of course, thanks to the \emph{reconstruction theorem} \cite{RT}, there is no loss of information as compared to a property which would bear directly on the generating functional itself, from which the $EL$ property can be obtained also \cite{prepara}.

 \subsection{Effective Locality at eikonal and quenched approximation. The gluon bundle}
The reduction formula being applied to (\ref{4pts}), in order to obtain an $S$--matrix element, one is led to the following eikonal and quenched scattering amplitude,
\begin{eqnarray}\label{4ptsF}
& & {\bf M}( p_1, p'_1; p_2, p'_2) \simeq {\bf M}( p_1; p_2) = \, \mathcal{N} \, \int\!\!{d[\chi] \, e\,^{\displaystyle{\frac{i}{4}\int{\chi^{2}}}} \, e\,^{\displaystyle{\mathfrak{D}_{A}^{(0)}}} \,}\, e\,^{\displaystyle{\frac{i}{2}\int{\chi\,F} -\frac{i}{2}\int{A\, \partial^2 A }} }\nonumber\\  &\times & G_{\rm eik}^1(p_1|A) \, G_{\rm eik}^2(p_2|A)\,\Bigl|_{\displaystyle{A=0}}
\end{eqnarray}
describing the collision of two quarks with their respective 4--momenta $p_1$ and $p_2$ being unchanged before and after the collision, $p'_1\simeq p_1; p'_2\simeq p_2$, neglecting the spin effects ( the eikonal approximation~) and the fermionic loop $L[A]$ being set equal to $0$ ( the quenched approximation ).

In this frame, the last line of eq.(\ref{Fradkin}) simplifies to (\cite{Herb} chap.7\&8, \cite{Eik} chap.5) :
\begin{equation}\label{Geik}G_{\rm eik}(p|A)\propto T_s\,e\,^{\displaystyle{-ig\,p^\mu\!\int_{-\infty}^{+\infty}\!\! {\rm{d}}s\,A^a_\mu(y-sp)\,{\lambda^a\over2}}}\end{equation}

 The $A^a_\mu$ field of (\ref{Geik}) belongs to a time--ordered exponential. This difficulty can be dealt with by introducing extra field variables, writing for example, with $\mathcal{N}$, another constant of normalization ,
\begin{eqnarray}\label{out}
&{}&\nonumber T_s\,e\,^{\displaystyle{-ig\,p^\mu\!\int_{-\infty}^{+\infty}\!\! {\rm{d}}s\,A^a_\mu(y-sp)\,{\lambda^a\over2}}}= \displaystyle\mathcal{N}\int\!\! {\rm{d}}[\alpha]\int\!\!{\rm{d}}[\Omega]\, e\,^{\displaystyle{i\!\int_{-\infty}^{+\infty}\!\! {\rm{d}}s\,\,\Omega^a(s)\,\alpha^a(s)}}\nonumber\\  &{}&\times\, e\,^{\displaystyle{-ig\,p^\mu\!\int_{-\infty}^{+\infty}\!\! {\rm{d}}s\,\,\Omega^a(s)\,A^a_\mu(y-sp)}} T_s\,e\,^{\displaystyle{-i\int_{-\infty}^{+\infty}\!\!{\rm{d}}s\,\alpha^a(s)\,{\lambda^a\over2}}}
\end{eqnarray}

so that finally, one can write for the exponential involving the $A_{\mu}^a$ field, 
\begin{equation}\label{R} e\,^{\displaystyle{-ig\,p^\mu\!\int_{-\infty}^{+\infty}\!\! {\rm{d}}s\,\,\Omega^a(s)\,A^a_\mu(y-sp)}} = e\,^{\displaystyle{-i\int_{-\infty}^{+\infty}\!\! {\rm{d}}^4z\,\,R_{\mu}^a(z)\,A^a_\mu(z)}}\end{equation}
with,
 \begin{equation}\label{QC}\displaystyle R^a_{\mu}(z) = g\,p_{\mu}\, \int_{-\infty}^{+\infty}\mathrm{d}s\,\Omega^a(s)\,\delta^{(4)}(z - y + s\, p)\end{equation}

In this quark--current expression (\ref{QC}), the eikonal approximation for the Fradkin field variable $u_{\mu}(s)$ consists of a straight line relation connecting the points $z$ and $y$, $u_{\mu}(s) = s\,p_{\mu}$.

The purpose of expressions (\ref{Geik}--\ref{QC}) is to offer as simple as possible a derivation of the $EL$ property. As exact integrations on the auxiliary field variables $\alpha^a$ and $\Omega^a$ and on $z$ are achieved, one is insured to deal with the proper Fradkin representation of $G_F[A]$ in the eikonal approximation physically relevant to the $EL$ regime. 

\par
The result obtained in this way, together with the quenched approximation $L[A] = 0$,  has the following form  :
\begin{eqnarray}\label{ELEQ}
&&{\bf M}( p_1; p_2) = \displaystyle\,\prod_{i=1}^{2}\int\mathrm{d}u_i(s_i)\int\mathrm{d}\alpha_i(s_i)\int\mathrm{d}\Omega_i(s_i)\,\left(\ldots\right)\\ &\displaystyle\times& \nonumber \int\!\mathrm{d}[\chi] \, e\,^{\displaystyle{ \frac{i}{4} \int{ \chi^{2} }} } \, e\,^{\displaystyle{\mathfrak{D}_{A}^{(0)}}} \, e\,^{\displaystyle{\frac{i}{2} \int\!{ A ^a_\mu\, K^{\mu\nu}_{ab}\, A^b_\nu} }} \, e\,^{\displaystyle -i\int\!{Q^a_\mu A^\mu_a } }\,\bigg|_{\displaystyle A = 0 }
\end{eqnarray}
where $K^{\mu\nu}_{ab}$ and $Q^a_\mu$ factorize the quadratic and linear $A^a_\mu$ field dependences respectively :
\begin{equation}\label{QK}
K_{\mu\nu}^{ab}=gf^{abc}\chi_{\mu\nu}^c- g_{\mu\nu}\,\delta^{ab}\,\partial^2 = (gf\!\cdot\! \chi)^{ab}_{\mu\nu}- (D_{\mathrm{F}}^{(0)-1})_{\mu\nu}^{ab}\,, \ \ \ Q^a_\mu =\partial^\nu\chi^a_{\nu\mu}+ (R^a_{1,\mu}+R^a_{2,\mu})\end{equation}
and where the $R^a_{i,\mu}$ are the leading parts of $Q^a_\mu$ in the strong coupling regime, $g\!\gg\!1$.
The linkage operation can operate easily now, and setting the sources $j^a_\mu$s to zero afterwards yields :
\begin{eqnarray}\label{elform}
& & e\,^{\displaystyle{\frac{i}{2} \int{\frac{\delta}{\delta A} \,{D}_{F}^{(0)} \, \frac{\delta}{\delta A} }}} \,\,e\,^{\displaystyle{\frac{i}{2} \int\!A \,K \,A - i \int \!Q\,A }}\,\biggr|_{A \rightarrow 0} \\ \nonumber &=&  \,\,e\,^{\displaystyle-{\frac{i}{2} \int{{Q}\left( {D}_{\mathrm{F}}^{(0)} \,\left( 1 + {K} \, {D}_{\mathrm{F}}^{(0)}\right)^{\!-\!1} \right){Q}}}}\,e^{\displaystyle{-\frac{1}{2} \Tr{\ln{\left( 1+ {D}_{\mathrm{F}}^{(0)} \,{K} \right)}}}}
\end{eqnarray}

The kernel of the quadratic term in ${Q}_{\mu}^{a}$ in the right hand side of (\ref{elform}) is :
\begin{eqnarray}\label{magics}
{D}_{\mathrm{F}}^{(0)} \,\left( 1 + {K} \, {D}_{\mathrm{F}}^{(0)}\right)^{\!-\!1} &=& {D}_{\mathrm{F}}^{(0)} \,\left( 1 + \left( g f \!\cdot\! \chi - {{D}_{\mathrm{F}}^{(0)}}^{\!-\!1} \right) \, {D}_{\mathrm{F}}^{(0)}\right)^{\!-\!1} \\ \nonumber &=&  \left( g f \!\cdot\!\chi \right)^{\!-\!1}
\end{eqnarray}
 and in all, the last line of (\ref{ELEQ}) reads :
\begin{eqnarray}\label{EL}
& & \int{\mathrm{d}[\chi] \, e\,^{\displaystyle{ \frac{i}{4} \int{ \chi^{2} }} }}\,\ e\,^{\displaystyle{\frac{i}{2} \int{\frac{\delta}{\delta A} \, {D}_{\mathrm{F}}^{(0)} \, \frac{\delta}{\delta A} }} }\,\,e\,^{\displaystyle{ \frac{i}{2} \int{A \,{K} \, A} - i \int \!Q\,A} }\, \biggr|_{\displaystyle A = 0} \\ \nonumber &=& e\,^{\displaystyle{-\frac{1}{2} \Tr{\ln{\bigl(g{D}_{\mathrm{F}}^{(0)}\bigr)}}}}\int\frac{\mathrm{d}[\chi]}{\sqrt{\det(f\!\cdot\!\chi)}} \,\, e\,^{\displaystyle{ \frac{i}{4} \int{ \chi^{2} }} }}\,e\,^{\displaystyle{-\frac{i}{2} \int\mathrm{d}^4z\ {{Q}(z) \,(gf\!\cdot\!\chi)^{-1}(z)\, {Q}(z)}}
\end{eqnarray}

On the right hand side, the first term is a (possibly infinite) constant to be absorbed into a redefinition of the overall normalization $\mathcal{N}$, and it is in the last term of (\ref{EL}) that the Effective Locality phenomenon is finally seen. In effect, the ${D}_{\mathrm{F}}^{(0)}$--pieces entering (\ref{QK}), (\ref{elform}) and (\ref{magics}) are non local but disappear from the final result leaving a structure which turns out to be \emph{local}, $\langle z | (gf\!\cdot\!\chi)^{-1} | z' \rangle = (gf\!\cdot\!\chi)^{-1}(z) \, \delta^{(4)}(z-z')$, and which mediates an \emph{effectively local interaction} between `` gluonic ''and quark currents, in the form : ${Q}(z) \,(gf\!\cdot\!\chi)^{-1}(z)\, {Q}(z)$. 

\begin{figure}
\centering
\includegraphics[width=5cm]{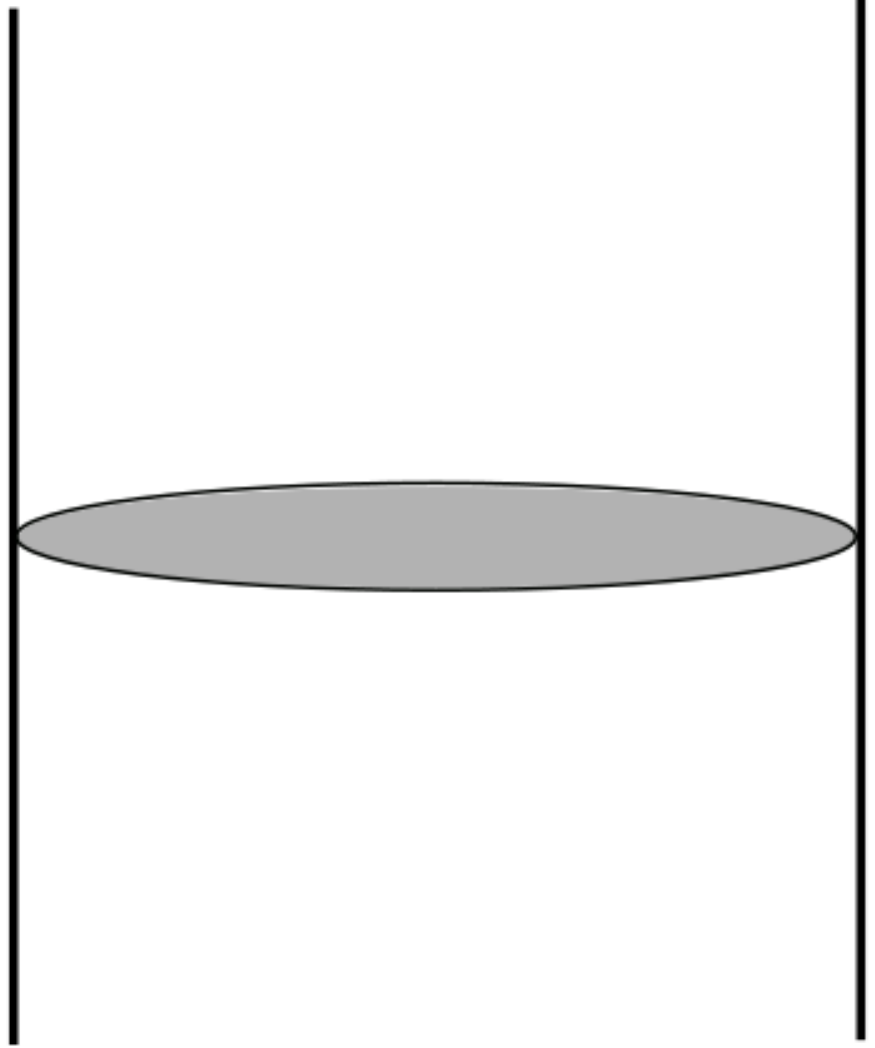}
\caption{A gluon bundle ( shaded oval ) exchanged between two quarks (solid lines).}
\label{quark_gluonbundle}      
\end{figure}

This $(gf\!\cdot\!\chi)^{-1}(z)$ term stands now for the original gluon propagator ${D}_{\mathrm{F}}(z - z')$ and has been called a \emph{gluon bundle} \cite{QCD-II}. The $R_1(z) \,(gf\!\cdot\!\chi)^{-1}(z)\, R_2(z)$ amplitude can accordingly be represented by the diagram of Fig.\ref{quark_gluonbundle}. A convenient expression deduced from ${Q}(z) \,(gf\!\cdot\!\chi)^{-1}(z)\, {Q}(z)$, and its relation to Fig.1,  is given in eq.(\ref{Xsq2}) and its subsequent paragraph.
\par
To sum up, equations (\ref{ELEQ}) to (\ref{EL}) display the $EL$ property at quenched and eikonal approximation.

\subsection{Effective Locality as general, still formal a statement}
Remarkably enough, the $EL$ property extends to the full non approximate $QCD$ theory \cite{QCD-II}, at least in a formal way. On the basis of the same $4$--point function, the expression equivalent to (\ref{EL}) reads in this case :
\begin{eqnarray}\label{surprising}
R.H.S \,(\ref{EL})\,\longrightarrow & & e\,^{\displaystyle{-\frac{1}{2} \Tr{\ln{\bigl({D}_{\mathrm{F}}^{(0)}\bigr)}}}} \int{\mathrm{d}[\chi] \, e\,^{\displaystyle{ \frac{i}{4} \int{ \chi^{2} }} }} \,e\,^{\displaystyle{-\frac{i}{2} \int{\bar{{Q}} \,\widehat{{K}}^{-1} \, \bar{{Q}}} - \frac{1}{2} \Tr\ln{\widehat{{K}}} }}\nonumber \\  & & \times\,e\,^{\displaystyle{\frac{i}{2} \int{\frac{\delta}{\delta A} \, \widehat{{K}}^{-1} \, \frac{\delta}{\delta A}} + \int{\bar{{Q}} \, \widehat{{K}}^{-1} \, \frac{\delta}{\delta A} }}} \,e\,^{\displaystyle{ {{L}}[A] }} \biggr|_{A\rightarrow 0}
\end{eqnarray}with the new definitions of `kernel' and `currents' :
\begin{eqnarray}\label{fullK}
\langle z | \widehat{{K}}_{\mu \nu}^{ab} |z' \rangle = \bigl(\, {{K}}_{\mu \nu}^{ab}(z) + g f^{abc} \chi_{\mu \nu}^{c}(z)\, \bigr) \, \delta^{(4)}(z - z')
\end{eqnarray}
\begin{eqnarray}\label{22}
\bar{{Q}}_{\mu}^{a}(z) = & &\partial^{\nu} \chi_{\nu \mu}^{a}(z) + 2g\, \partial^{\nu} \Phi_{\mathrm{I}, \nu \mu}^{a}(z) + g \int_{0}^{s_{1}}{\mathrm{d}s_1' \, \delta^{(4)}(z - y_{1} +u_1(s_1')) \, u'_{1,\mu}(s_1')\, \Omega_{\mathrm{1}}^{a}(s_1')} \nonumber\\  & & + \,2g\, \partial^{\nu} \Phi_{\mathrm{2},\nu \mu}^{a}(z) + g \int_{0}^{s_{2}}{\mathrm{d}s_2' \, \delta^{(4)}(z - y_{2} + {u_2}(s_2')) \, u'_{2,\mu}(s_2')\, \Omega_{\mathrm{2}}^{a}(s_2')}
\end{eqnarray}

Previously absent from the eikonal and quenched case, spin contributions show up now, and in (\ref{22}), the new spin--related terms read :
\begin{eqnarray}\label{Eq:23}
\Phi_{i,\mu \nu}^{a}(z;y_i) \equiv\Phi_{i,\mu \nu}^{a}(z) \equiv \int_{0}^{s_i}{\mathrm{d}s_i' \, \delta^{(4)}(z - y_i +u_i(s_i')) \, \Phi_{i,\mu \nu}^{a}(s_i')},\ \ \ \ i=1,2
\end{eqnarray}
while, for the kernel, the new spin--related term is :
\begin{eqnarray}\label{spins}
{{K}}_{\mu \nu}^{ab}(z) = & & 2 g^{2} \int_{0}^{s_{1}}{\mathrm{d}s_1' \ \delta^{(4)}(z - y_{1} + u_1(s_1')) f^{abc}\, \Phi_{\mathrm{1},\mu\nu}^{c}(s_1')}\nonumber \\  & & \!\!\!\!\!+\, 2 g^{2} \int_{0}^{s_{2}}{\mathrm{d}s_2' \ \delta^{(4)}(z - y_{2} + {u_2}(s_2')) f^{abc}\, \Phi_{\mathrm{2},\mu\nu}^{c}(s_2')}
\end{eqnarray}

Because of spins, in effect, extra fields, $\Phi^a_{1,\mu \nu}(s_{1})$ and $\Phi^a_{2,\mu \nu}(s_{2})$ are necessary to extract the $A_{\mu}^{a}(y - u(s'))$ potentials from under the full ordered exponential of (\ref{Fradkin}). This can be achieved by introducing extra functional integrations :
\begin{eqnarray}\label{full}
& & T_{s'}\,e\,^{\displaystyle{ -ig \int_{0}^{s_i}{ds_i' \, {u'}_i^{\mu}(s_i') \, A_{\mu}^{a}(y_i-u_i(s_i')) \, {\lambda^{a}\over2}} + ig \int_{0}^{s_i}{ds_i'\, \sigma^{\mu \nu} \, {F}_{\mu \nu}^{a}(y_i-u_i(s_i')) \, {\lambda^{a}\over2}}} }\nonumber \\ \nonumber &=& \mathcal{N}_{\Omega} \, \mathcal{N}_{\Phi} \, \int d[\alpha_i^a] \, \int d[\Xi_{i\mu\nu}^a] \,  T_{s'}\,e\,^{\displaystyle{ -i \int_{0}^{s_i}{ds_i' \, \left( \alpha_i^{a}(s_i') -  \sigma^{\mu \nu} \, {\Xi_i}_{\mu \nu}^{a}(s_i') \right) \, {\lambda^{a}\over2}}}  } \\ & & \times\, \int d[\Omega_i^a] \, \int d[\Phi_{i\mu\nu}^a] \,e\,^{\displaystyle{i \int_0^{s_i}{ds_i' \, \Omega_i^{a}(s_i') \, \alpha_i^{a}(s_i')}  - i \int_0^{s_i}{ds_i' \, {\Phi_i}^{a\mu \nu}(s_i') \,  {\Xi_i}^{a}_{\mu \nu}(s_i') } }} \\ \nonumber & &\times\, e\,^{\displaystyle{- i g \int_0^{s_i}{ds'_i \, u'^{\mu}(s_i') \, \Omega_i^{a}(s_i') \, A^{a}_{\mu}(y_i-u_i(s_i')) } + i g \int_0^{s_i}{ds_i' \, {\Phi_i}^{a}_{\mu \nu}(s_i') \, {F}^{\mu \nu}_{a}(y_i-u_i(s_i'))}  }}
\end{eqnarray}and as shown in the Appendix, the closed loop functional ${L}[A]$ can be endowed with similar functional integrations so as to take linear and quadratic $A^a_\mu$--gauge field dependences outside of the ordered exponentials. 
\par
Dropping ${L}[A]$, the second and fourth terms on the right hand sides of (\ref{22}), and proceeding to the eikonal replacement of $u_i(s_i)$ by $s_ip_i$, one immediately recovers the eikonal and quenched results of (\ref{QK}) to (\ref{EL}). In the case of higher number of points fermionic Green's functions, additional terms will of course complete the ${{\bar{Q}^a_\mu}}$ and ${\widehat{K}}_{\mu \nu}^{ab}$ functionals in exactly the same way \cite{QCD-II,RefI}.
\par\medskip
In (\ref{surprising}) nothing ever more refers to ${D}_{\mathrm{F}}^{(0)}$, that is to any initial gauge--fixing condition, as exactly the same result is reached whatever the gauge--field function in use, ${D}_{\mathrm{F}}^{(0)}$, ${D}_{\mathrm{F}}^{(\zeta)}$, ${D}_{\mathrm{F}}^{(\mathbf{n})}$, etc \cite{tg}. Assuming, as in the previous quenched situation, that integrating over the extra fields $\Xi$ and $\Phi$ can be carried through, this conclusion is non trivial.  \par
The striking aspect of (\ref{surprising}) is that, because the kernel $\widehat{{K}} = {{K}} + (g f\!\cdot \!\chi)$, so as the corresponding linear and quadratic parts coming from ${L}[A]$, are all local functions, with non zero matrix elements $\langle z | {{K}} | z' \rangle = {{K}}(z) \, \delta^{(4)}(z - z')$, the contributions of (\ref{surprising}) will depend on Fradkin, Halpern and subsidiary field variables in a specific, but \emph{local} way. 
\par
That is, again, the peculiar aspect of (\ref{surprising}) is its locality even though (\ref{surprising}) doesn't provide the bases of a genuine dual formulation of the full $QCD$ theory. In particular, it is easy to see that the fundamental duality correspondence of $g\rightarrow 1/g$ cannot account for the various $g$--scaling behaviors that can be exhibited out of (\ref{surprising}).
 
\section {\label{SEC:3}Theoretical aspects of effective locality}

First of all it is important to emphasize that the $EL$ property is non--perturbative. While for $QCD$ in its perturbative regime, gluons are pertinent dynamical degrees of freedom and are experimentally checked, in $EL$ calculations of fermionic Green's functions gluons disappear to the exclusive benefit of the $\chi^a_{\mu\nu}$--fields ${}^{\footnotemark[1]}$\footnotetext[1]{In Euclidean pure Yang--Mills theory, and at leading order of a semi classical expansion, the $\chi^a_{\mu\nu}$--fields are the usual strength field tensors evaluated over \emph{instanton field} configurations \cite{RefF}, \textit{i.e.}, $\chi^a_{\mu\nu}=F^a_{\mu\nu}(A^b_{inst.})$. }. This will be shown to be even more blatant in the pure Minkowskian Yang--Mills case \cite{prepara}.
\subsection{Fradkin's representation independence}
 In all these derivations, approximate or not, it would seem that the $EL$ property is heavily dependent on the Fradkin's expressions (\ref{Fradkin}) being used to represent the quark field propagators $G_F(x,y|A)$ in a given background gauge field.
\par
Now, it can be shown that provided the $\chi^a_{\mu\nu}$--field integrations are carried through, the $EL$ property turns out to be gauge invariant and independent of any representation used for the quark field functions $G_F(x,y|A)$ and would appear, thus, as a sound non--perturbative property of $QCD$ \cite{tg, prepara}.

\subsection{An odd term : { ${\bf\delta^{(2)}}(\vec{\bf b})$}}

Now, as Fradkin's representations (\ref{Fradkin}), which are exact, are implemented, then the $EL$ property comes along with a mass scale. For fermionic Green's functions to be sensitive a mass scale must be introduced. This fate is related to the occurrence of Dirac deltas in (\ref{Fradkin}) and (\ref{22}) : In the interaction term of (\ref{EL}), second line, they lead to a $\delta^{(2)}({\vec{b}})$, where $\vec{b}\equiv y_{1\perp}- y_{2\perp}$ is the \emph{impact parameter} of the two quark scattering process in their \emph{center of mass system}.
\par
It is important to stress that this $\delta^{(2)}({\vec b})$ term is not related to any sort of approximation. First discovered in an eikonal approximation giving the Fradkin fields $u(s)$ the simple straight line form of $u_i(s_i)=p_is_i$, it is possible to prove that in the much enlarged context of Fradkin's fields taken as elements of a \emph{Wiener functional space}, the same odd $\delta^{(2)}({\vec{b}})$ term comes about~\cite{QCD6}. 
\par
In Ref.\cite{QCD1} a physical argument has been suggested to account for this situation : This factor of $\delta^{(2)}({\vec{b}})$ may be a remnant of the implicit existence of asymptotic quark states, an assumption which, beyond the stage of perturbation theory cannot be maintained in $QCD$, neither theoretically nor experimentally \cite{Lavelle1996}, while it is contained in (\ref{Fradkin}) ${}^{\footnotemark[2]}$ \footnotetext[2]{This is because the $QCD$ Lagrangian is written out of the short distance dynamical degrees of freedom.}.
\par
In $QCD$ where both confinement and chiral symmetry breaking are realized, it is known that the inter--quark separation fluctuates and cannot be zero \cite{Stan}. That is, from this point of view also, the odd $\delta^{(2)}({\vec{b}})$ cannot make sense and must be modelled into a more pertinent impact parameter distribution. Compelling reasons exist to motivate the following substitution \cite{Matveev2002} :
\begin{equation}\label{deltaphy}
{\delta^{(2)}(\vec{b})}\ \longrightarrow\  \varphi(b)=\displaystyle\frac{\mu^{2}}{\pi} \, \frac{1 + \xi/2}{\displaystyle\Gamma(\frac{1}{1+\xi/2})} \, e^{\displaystyle-(\mu b)^{2+\xi}} , \quad \xi\in{\mathbb{R}},\ \  |\xi| \ll 1
\end{equation}where $\mu$ is the $EL$ property mass scale and $\xi$ a small \emph{deformation parameter}.  They both will be given a value in section \ref{SEC:4}.B 
\par
As illustrated, when it comes to chiral symmetry breaking, in a limit process, the $EL$ mass scale or equivalent distance will prevent the two points $x$ and $y$ to be taken arbitrarily close to each other, and, not to be lead back to the short distance behavior of $QCD$ where gluons, instead of $\chi^a_{\mu\nu}$--fields, are the relevant dynamical degrees of freedom. 

\subsection{Integration measures}
In (\ref{EL}) and (\ref{surprising}), a functional measure of integration remains, on $\chi^a_{\mu\nu}$--fields, and reads :
 \begin{equation}\label{init-m}
\int{\mathrm{d}[\chi]} = \prod_{w_i\in\mathcal{M}} \prod_{a=1}^{N_c^2-1} \prod_{ 0= \mu<\nu}^3 \int {\mathrm{d}[\chi_{\mu \nu}^{a}(w_{i})]}
\end{equation}but is here meant in a somewhat symbolic form so long as the measure on the infinite dimensional functional space of $\chi^a_{\mu\nu}$--configurations is not properly defined. In the very case of fermionic Green's functions, though, it turns out that, besides the already discussed $\delta^{(2)}(\vec{b})$, products of Dirac deltas single out a unique point of interaction $w_i$ in $\mathcal{M}$ \cite{QCD1,QCD6}; typically in the second line of (\ref{EL}) one would have ($w_i=y_i-u_i(s_i),\ i=1,2$) :
\begin{equation}\label{local}\int\mathrm{d}^4z\ {{R}(z;y_1,y_2) \!\cdot (gf\!\cdot\!\chi(z))^{-1}\cdot {R}(z;y_1,y_2)}\ \longrightarrow\  R(w_i) \cdot (gf\!\cdot\!\chi(w_i))^{-1}\cdot {R}(w_i)\end{equation}

This has the most favorable consequence that the \emph{measure image theorem}${}^{\footnotemark[3]}$ \cite{Ted}
\footnotetext[3]{When passing from an infinite dimensional functional space to a finite dimensional one (where Random Matrix Theory is used), this theorem can be viewed as generalizing the more customary notion of a \emph{Jacobian}.}can be invoked to translate the original and formal functional space measure (\ref{init-m}) into a well defined measure of integration on the finite dimensional space of \emph{skew--symmetric matrices} $\mathbb{M}\equiv -i\displaystyle\sum_{a=1}^{N_c^2-1}{{\chi^a_{\mu\nu}}\otimes T^a}$ :
\begin{eqnarray}\label{new-m}
-i\, {\rm{d}}(\sum_{a=1}^{N_c^2-1}{\chi^a}_{\mu \nu}\otimes T^a)&\equiv& {\rm{d}}\mathbb{M}= {\rm{d}}M_{11}\,{\rm{d}}M_{12} \cdots {\rm{d}}M_{NN} \nonumber \\ &=&\nonumber \left|\frac{ \partial(M_{11}, \cdots, M_{N\!N})}{\partial(\xi_1, \cdots, \xi_N, p_1, \cdots, p_{N(N-1)/2})}\right| \, {\rm{d}}\xi_1 \cdots {\rm{d}}\xi_N \, {\rm{d}}p_1 \cdots {\rm{d}}p_{N(N-1)/2} \\  &=& \prod_{i=1}^{N}\ {\rm{d}}\xi_i  \prod_{1\leq i<j}^N |\xi_i-\xi_j|^{\kappa}\   {\rm{d}}p_1\  ..\ {\rm{d}}p_{N(N-1)/2}\, f({\mathbf{p}})
\end{eqnarray}where the $\xi_i$s are the eigenvalues of $\mathbb{M}$ and where $\kappa=1$ and where the very last factors of (\ref{new-m}), define a \emph{Haar measure} of integration on the orthogonal group $O_N(\mathbb{R})$. 
\par
To sum up, the $EL$ property allows one to rely on the measure image theorem, to transform the initial functional measure of integration ${\mathrm{d}}[\chi]$ into the product of an integration on the spectrum of $\mathbb{M}$, ${\mathrm{Sp}}(\mathbb{M})$, times an integration on the orthogonal group $O_N(\mathbb{R})$; symbolically, one has :
\begin{equation}  
{\mathrm{d}} [\chi]\ \longrightarrow\ {\mathrm{d}}\,{\mathrm{Sp}}(\mathbb{M})\times {\mathrm{d}}\, O_N(\mathbb{R})\end{equation}

\subsection{Effective locality and \emph{Meijer} special functions}
In eikonal and quenched approximations, it is remarkable that the strong coupling fermionic $QCD$ Green’s functions can be calculated without further approximations. This is due to the peculiar $EL$ property (\ref{local}) and achieved by means of a standard analytical continuation of the \emph{Random Matrix} treatment \cite{QCD6}.
\par
Inserting (\ref{new-m}) in (\ref{EL}) and integrating on ${\mathrm{Sp}}(\mathbb{M})$ only, fermionic Green's functions show up as finite sums of finite products of \emph{Meijer} special functions \cite{GR}.  One finds effectively :
\begin{eqnarray}\label{Meijer}
&& \nonumber  \mathbf{M}(x_{1}, y_{1}; x_{2}, y_{2})=\mathcal{N}\,(-{16\pi^2 m^2\over E^2})^N\sum_{\mathrm{monomials\, \{q_1,\dots,q_i,\dots,q_N \}}}\biggl\langle\,\prod_{i=1}^N\, [1-i(-1)^{q_i}] \\ &&\nonumber\times\, \biggl({{\sqrt{2iN_c}}\,{\sqrt{{\widehat{s}}({\widehat{s}}-4m^2)}}\over {m^2}}\biggr)\, \frac{(({\cal{OT}})_i)^{-2}}{g\varphi(b)}\\ &&\times\, G^{30}_{03}\!\left( \biggl({  g\varphi(b)\over {\sqrt{32iN_c}}  }{m^2\over {\sqrt{{\widehat{s}}({\widehat{s}}-4m^2)}}}\biggr)^2\bigl(({\cal{OT}})_i\bigr)^{4}\, \biggr|\frac{1}{2}, \frac{3+2q_i}{4},1\!\right)\,\biggr\rangle_{\!O_N(\mathbb{R})}
\end{eqnarray}in the same example of a $4$--point Green's function such as (\ref{4pts}), and where $G^{30}_{03}$ is the \emph{Meijer} special function \cite{pomme}
\begin{eqnarray}\label{Meijer2}
\int_0^\infty\ {{\rm{d}}\xi}\ \ \xi^p\ \,e^{\displaystyle{-\xi^2-{b\over \xi}}}}={\frac{1}{2 \sqrt{\pi}}\, G^{30}_{03}\left(\frac{b^2}{4} \biggr| {p+1\over 2}, \frac{1}{2},0\right)
\end{eqnarray}

Equality holds for $p>0$ and $b>0$, whereas the point is that the right hand side, the \emph{Meijer} function, is analytic in its argument, $b^2/4$, \cite{pomme}. In ({\ref{Meijer}) one has also $N\equiv D\times(N_c^2-1)=32$, at $D=4$ spacetime dimensions, while $\hat{s}=(p_1+p_2)^2$ and $E=\hat{s}/2$ in the center of mass system of the scattering quarks of $4$--momenta $p_1$ and $p_2$.
\par
In the expression (\ref{Meijer}), ${\cal{O}}$ is an orthogonal matrix and integration on $O_N(\mathbb{R})$ remains to be done, as meant by the overall average's brackets.

\par Letting aside momentarilly the various elements entering (\ref{Meijer}), the point here is that, in full generality, Green's functions and/or their generating functionals have been proven to be expressible in terms of \emph{Fox} special functions, which are but mere generalisations of \emph{Meijer}'s special functions. 
\par
Within the above approximations at least, $EL$ calculations can be seen to comply with this general statement \cite{Ferrante}.

\subsection{Color algebraic structure of fermionic Green's functions}
 As observed in non relativistic quark models \cite{Dmitrasinovic} and in a non abelian generalization of the Schwinger mechanism \cite{Nieuwenhuizen}, an additional dependence on the $SU_c(3)$ cubic Casimir operator $C_{3f}$ shows up, in contradistinction with perturbation theory and other non--perturbative
approaches, which only display quadratic Casimir operator, $C_{2f}$ dependences. Though numerically sub--leading, these extra $C_{3f}$ dependences account for the full algebraic content of the rank--2 Lie algebra of $SU_c(3)$ and there are \emph{a priori} no super selection arguments to discard them.  
\par
With $z_i$ the argument of $G^{30}_{03}$ in (\ref{Meijer}), one has $z_i=\lambda\, \bigl(({\cal{OT}})_i\bigr)^{4}$ where ${\cal{T}}$ is the $32$--vector of components made out of $4$ copies of the full set of Gell--Mann generators, the $\{\lambda^a/2\}$. The parameter $\lambda$ can be read off (\ref{Meijer}) and turns out to be a very small parameter even at large enough coupling constant $g\sim 15$, so that an analytic expansion of $G^{30}_{03}$ can be devised. The average on the orthogonal group $O_N(\mathbb{R})$ can be carried out to get a leading contribution of :
 \begin{equation}\label{leading}\langle{\sqrt{z_i}}\rangle_{O_N(\mathbb{R})}\,=\frac{\sqrt{\lambda}}{N}\,DC_{2f}\, {\mathbf{1}}_{3\times 3}\end{equation}and on the fundamental representation the quadratic Casimir eigenvalue is $C_{2f}=4/3$

At next to leading order now, the result reads \cite{tg} :
\begin{equation}\label{C3}
\langle z_i\rangle_{O_N(\mathbb{R})}\,=(\frac{\sqrt{\lambda}}{N})^2\left(\,(DC_{2f})^2+(DC_{3f})\right){\mathbf{1}}_{3\times 3}
\end{equation}and an extra $C_{3f}$ dependence shows up which extends presumably to all of the next to leading orders; for example :
\begin{equation}\label{C3bis}
\langle z_i\sqrt{z_i}\rangle_{O_N(\mathbb{R})}\,=(\frac{\sqrt{\lambda}}{N})^3\left(\,(2+(\frac{5}{6})^2)(DC_{2f})^2+(DC_{2f})(DC_{3f})+{3}(DC_{3f})\right){\mathbf{1}}_{3\times 3}
\end{equation}

These averages are independent of the index $i=1,2,\dots,32$. On the fundamental representation where it must be evaluated the cubic Casimir operator has eigenvalue $C_{3f}=10/9$. Numerically, the result of (\ref{C3}) shows that at order $({\sqrt{\lambda}/N})^2$, the trilinear Casimir operator $C_{3f}$ enhances the pure $C_{2f}$ contribution a non negligible amount of $ 15.6\%$, whereas at sub--leading order $({\sqrt{\lambda}/N})^3$, $C_{2f}$ and $C_{3f}$ contributions to (\ref{C3bis}) are identical to within $.2\%$.

\par
Here and elsewhere \cite{Nieuwenhuizen, Dmitrasinovic, Nayak}, $C_{3f}$ dependences may be sub--leading effects but are proposed to be viewed as hallmarks of the $QCD$ non--perturbative fermionic sector. Within the above approximations at least, $C_{3f}$ dependences are clear outputs of $EL$ calculations.

\subsection{$EL$ calculations : non--perturbative and gauge invariant}
 $EL$ calculations are gauge invariant \cite{tg, Rhodos,  prepara}. Attempts at trading an $A^a_\mu$ potential formulation of $QCD$ for an  $F^a_{\mu\nu}$ field strength formulation exist already \cite{RefF}, relying on the same `linearization' trick as (\ref{chi}). In the quantization process though, covariant gauge conditions were chosen, ultimately transferred to the $\chi^a_{\mu\nu}$--field. That is, in the non--perturbative strong coupling/strong field limits, the famous \emph{Gribov copy} problem is met again with, so far, no envisageable solutions and no sound control of gauge invariance.
\par
The reason why $EL$ calculations escape this dead end is that in the $EL$ context, quantisation is achieved by \emph{functional differentiations}, with the help of (\ref{link}), rather than functional integrations with gauge-fixing terms. Both quantisations are equivalent whenever the \emph{Wick theorem} holds true for time ordered products of field operators \cite{Kleinert}.
\par
Now, as written in the Introduction for the $EL$ statement, and as can be stared at in (\ref{magics}) and (\ref{EL}), no gauge condition has ever been fixed in an $EL$ calculation, and to give things the more concise of all possible expression, it must be stated that there is no Gribov copy to cope with because there is no gauge--fixing condition to be copied. 
\par
By adding and subtracting adequate Lagrangian densities, any gauge--free field propagator can be generated, \emph{Feynman, general covariant, axial planar}, $D_F^{(0)}$, $D_F^{(\zeta)}$, $D_F^{(n)}$, etc ${}^{\footnotemark[4]}$...\footnotetext[4]{Including forms which would correspond to any choice of non linear gauge fixing conditions.} This is a mandatory step in order to make well defined an intermediate step ${}^{\footnotemark[5]}$ of an $EL$ calculation\footnotetext[5]{That is, inverting a previously non invertible quadratic form on the $A^a_\mu$ fields.}. Though this procedure generates, by construction, any possible gauge field propagator, it is important to realise that the propagators so obtained are not related to the insertions of corresponding gauge fixing conditions, $\delta\left({\mathcal{F}}[A^a_\mu]\right)$, in the context of functional integration; and that the theory quantized in this way~${}^{\footnotemark[6]}$ preserves, thereby, its original full gauge invariance. \footnotetext[6]{The details of this quantization are given in \cite{QCD1}.} 

\subsection{An effective perturbative expansion for the strong coupling regime}
Renormalization has been forced upon $QFT$s by the short distance \emph{ultra--violet} singularities which are met in perturbative expansions. Since $EL$ is \textit{a priori} relevant to larger distances, ultra--violet singularities should not be met in $EL$ calculations.
\par
This expectation is correct at quenched approximation, taking the closed quark loop functional $L[A]$ to zero. In this case in effect, the effective perturbative expansions defined out of the \emph{Meijer} function analyticity can be checked to be non trivial and to entail neither infrared nor any ultra--violet singularities \cite{prepar}. 
\par
However, as soon as the quenched approximation is relaxed and the fermionic loop functional is restored :
\begin{eqnarray}\label{Eq:LFradkin01}
{L}[A] &=&  - \frac{1}{2} \displaystyle\int_{0}^{\infty}{\frac{ds}{s} \, e^{\displaystyle-is m^{2}}} \, \mathcal{N}\,\int{d[u]} \, \delta^{(4)}(u(s)) \, e^{\, -\displaystyle\frac{i}{4} \int_{0}^{s}{ds' \, (u'(s'))^{2} } } \nonumber\\  & & \hskip-1truecm \times \int{d^{4}y \, \tr{\biggl( T_{s'}\,e\,^{\displaystyle -ig\!\int_0^s\!\!ds'\, u_{\mu}'(s')\,A_{\mu}^a(y-u(s')){\lambda^{a}\over2}+ ig\!\int_0^s\!\!ds'\,\sigma_{\mu\nu}\,F_{\mu\nu}^a(y-u(s')){\lambda^{a}\over2}}\, \biggr)} }\nonumber \\  & & - \left\{ g = 0 \right\}
\end{eqnarray}then usual ultra--violet singularities show up the usual way by the $s$ lower integration boundary, at $s= 0$. This is because in the closed quark loop functional $L[A]$, the $EL$ mass scale $\mu$ doesn't show up to prevent short distances contributions. 
\par
For example this is obtained out of the second line of (\ref{surprising}), as the functional translation operator has replaced the $A$ field by the $-\bar{{Q}} \cdot \widehat{{K}}^{-1}$-- combination, after the prescription of sending $A$ to zero is taken into account. At order $g^2$ of an expansion of $L[-\bar{{Q}} \cdot \widehat{{K}}^{-1}]$, one finds expressions like \cite{QCD5'} :
\begin{eqnarray}\label{ren}
g^{2}(q^{2};\Lambda^2) = g^{2} \,  \ln{\left( \frac{\Lambda^{2}} {m^{2} + q^{2} \, |{z}_{\mathrm{I} \, \mathrm{I\!I}}^{}| \, (1 - |{z}_{\mathrm{I} \, \mathrm{I\!I}}^{}|)} \right)}
\end{eqnarray}where the variable ${z}_{\mathrm{I} \, \mathrm{I\!I}}^{} \equiv {z}_{\mathrm{I}}^{} - {z}_{\mathrm{I\!I}}^{}$, is introduced and where $0\leq {z}_{\mathrm{I}},\,{z}_{\mathrm{I\!I}} \leq 1$, whereas $\Lambda$ stands for an ultra--violet cut--off. This expression displays the customary steps and features of the renormalization procedure, including a form of \emph{asymptotic freedom} appropriate to the $EL$ context \cite{QCD5'}. This should not come as too big a surprise, as asymptotic freedom has been recognised to extend beyond the very perturbative regime of $QCD$ \cite{Trento}.

\par
For the purpose of phenomenological applications relying on the eikonal and quenched approximations, this has the interesting consequence that the Lagrangian density one should  start from is the renormalised one \cite{QCD5'}, with the most favorable circumstance also, that no infrared nor any ultra--violet singularities will be met in the course of $EL$--perturbative expansions of fermionic amplitudes \cite{prepar}.

\subsection{$EL$ and dynamical chiral symmetry breaking}
Effective locality could thus exhibit the very way non--abelian gauge invariance is realized in the non--perturbative regime of $QCD$, avoiding the intractable and never ending issue of Gribov's copies \cite{Thess}. Now, if this property is really relevant to $QCD$, it should also shed some light on the fundamental issue of dynamical chiral symmetry breaking. 
\par
Calculations are involved and to simplify them somewhat, the quenched approximation is used : while it modifies the effect's magnitude, both theoretical and numerical analyses have long shown that the quenched approximation preserves the dynamical chiral symmetry breaking phenomenon, if any \cite{Dyakonov}. An eikonal approximation is used also, but in a \emph{mild} way so as to preserve chirality ${}^{\footnotemark[7]}$ and in order to allow for controlled calculations. 
\footnotetext[7]{A strict eikonal approximation would devoid chirality of any meaning.} 
\par
In the chiral limit of $m\rightarrow 0$ where the phenomenon is non--trivial, the order parameter of the chiral symmetry, $<\bar{\Psi}(x)\Psi (x)>$, can be obtained out of $<\bar{\Psi}(x)\Psi (y)>$ in the limit of $x\rightarrow y$ \cite{tgpt} :
\begin{eqnarray}\label{0x0}
<\bar{\Psi}(x)\Psi (x)>\,&=&\lim_{x\rightarrow y} \ i\mathcal{N}\,\delta^{(4)}(x-y)\ \Tr\!\int\! {\rm{d}}[\hat{\alpha}]\,e^{\displaystyle-i\frac{\delta s}{2}\,\hat{\alpha}\!\cdot\!\hat{\lambda} } \!\int\!{\rm{d}}[V]\, e^{\displaystyle-i \frac{\delta s}{4\Sigma}\,\, \hat{\alpha}\cdot V} \nonumber\\ 
 && \hskip-1truecm N^i\frac{\delta}{\delta V^i}\int \mathrm{d}\mathbb{M} \,\, \frac{e^{\,\displaystyle\frac{i}{8N_c}\Tr \mathbb{M}^2(y)}}{\sqrt{\det(\mathbb{M}(y))}}\,\,e^{\displaystyle-\frac{i}{2}\, g\,\frac{\mu^2\sqrt{\Delta}}{\pi Ep}\,
V\!\cdot\!\mathbb{M}^{-1}(y)\!\cdot\!V} \end{eqnarray}

However this expression is \emph{zero} for two reasons : once for a purely algebraic fact, similar to the algebraic identity $\Tr \gamma_\mu=0$, while a second `trivialization' occurs as the average on $O_N(\mathbb{R})$ is performed. 
\par
Fortunately, copying the massive $QED_2$ case \cite{FGHF}, both trivializations are circumvented by one and the same procedure which consists in calculating the appropriate $x_1, y_1, x_2, y_2$ limits of a $4$--point calculation ${}^{\footnotemark[8]}$. \footnotetext[8]{Contrary to the massive $QED_2$ case in effect, in $QCD$ one cannot rely on the property of \emph{cluster decomposition} \cite{tgpt}, taking the limits of $x_1(=y_1)$ and $x_2(=y_2)$ separated by an infinite spatial distance.}For the $2$--point function, one gets :
\begin{equation}\label{triv} \frac{C^{st} }{N}\left({0\,\times\,0}\right)\end{equation}while the relevant $4$--point result reads,
\begin{equation}\label{4}(\frac{C^{st}}{N})^2\left({C^{st}}'\neq 0\right)\end{equation}allowing to identify the order parameter of the chiral symmetry as $\langle\bar{\Psi}\Psi (x)\rangle_{partial}\,\simeq\, {C^{st}/ N}$,

where the sign of an approximate equality is to remind that approximations have been used. The subscript \emph{partial} means that (\ref{triv}) and (\ref{4})  are the contribution to $\langle\bar{\Psi}\Psi (x)\rangle$ of a given monomial among $2^{120}$ possible ones, alternate in signs, and coming from the Vandermonde determinant expansion of (\ref{new-m}). Accordingly, the full result appears unattainable.
\par
Fortunately again, Wigner's \emph{semi--circle law} can be used to circumvent the intractable task of evaluating the sum of so many monomials \cite{Mehta}. Within the following standard definitions of Random Matrix theory \cite{Mehta},
  \begin{equation}\label{M1}
P_{N\kappa}(\xi_1,\dots,\xi_N)\equiv \ C_{N\kappa}\prod_{ i<j}^N|\xi_l-\xi_j|^\kappa \,e^{\,\displaystyle{-}\sum_1^N\xi_i^2}\end{equation}
 \begin{equation}\label{M2}
 \biggl(\,\prod_{j=2}^N\int_{-\infty}^{+\infty}\,{\mathrm{d}\xi_j}\biggr)  \ P_{N2}(\xi_1,\dots,\xi_N)\equiv N^{-1}\,\sigma_N(\xi_1)\end{equation}the large $N$ limit of $\sigma_N(\xi)$ turns out to be simply given by :
  \begin{equation}\label{M3}
\sigma_N(\xi)\longrightarrow \sqrt{2N-\xi^2}\,, \ \ \mathrm{for}\  -\sqrt{2N}\leq \xi\leq +\sqrt{2N}\,,\ \  \sigma_N(\xi)=0\ \ \mathrm{otherwise}\end{equation}while sub--leading corrections can be calculated in a systematic way. From (\ref{M3}), long but rigorous calculations can be carried out and the chiral symmetry order parameter results as~:
\begin{equation}\label{condens1} 
    \lim_{y=x}\,\langle\bar{\Psi}(x)\Psi(y)\rangle\,\simeq -\,g^2\,\mu^3\,\frac{\mu}{\sqrt{Ep}}\,{\sqrt{\frac{E^2-p^2}{Ep}}}^{\,\,3}\,\frac{4^5(N_c-1)}{\sqrt{\pi^5N^3}} \,\frac{I(N)}{vol(O_N(\mathbb{R})}
\end{equation}where ${vol(O_N(\mathbb{R})}$ is the volume of the orthogonal group $O_N(\mathbb{R})$, and :
\begin{equation}\label{I1}
I(N)=\int_{-\sqrt{2N}}^{+\sqrt{2N}}\,{\mathrm{d}\xi\over \xi}\, {\sqrt{2N-\xi^2}}\ \,\Phi(\xi\sqrt{N})
\end{equation} 
entails the \emph{probability integral} $\Phi(x)$ \cite{GR}.
\par
Thus, most importantly, $EL$ involves dynamical chiral symmetry breaking out of $QCD$ first principles, and the chiral condensate goes like the third power of the Effective Locality mass scale $\mu$. 
\par
Then, appears something like \emph{a partonic depleting function}, here expressed in the center of mass system of a two quark scattering processs, \textit{i.e.}, identified out of a $4$--point Green's function \cite{tgpt, QCD6}, the function :
$$f(E, p;\mu)\equiv\frac{\mu}{\sqrt{Ep}}\,{\sqrt{\frac{E^2-p^2}{Ep}}}^{\,\,3}$$

If $E$ and $p$ are in a range of magnitude corresponding to the perturbative regime of $QCD$, \textit{i.e.}, $E,p>\Lambda_{QCD}$, then the particle--like character of quarks would allow $E^2-p^2$ to be replaced by $m^2$, the squared quark mass. But the interesting point is that the chiral condensate magnitude is modulated by this non--trivial function of quarks energy and momentum variables $E$ and $p$, in such a way that as $E$ and $p$ increase, the chiral condensate magnitude is depleted down to zero. This is in agreement with the fact that chiral symmetry breaking cannot be the result of a perturbative mechanism. Not innocuous also, the necessity of a $4$--point calculation : this could be related to physical peculiarities of the non--perturbative fermionic sector of $QCD$ \cite{tgpt}.


\section {\label{SEC:4}Phenomenological applications}

In this section, one goes through a number of phenomenological predictions obtained by using formula (\ref{EL}) and its generalisations deduced from (\ref{surprising}).

\subsection{Quark--quark binding potential}
In the eikonal approximation, the quark--quark scattering amplitude reads (\cite{Eik}, chap.8), 
\begin{equation}\label{Tsq} 
\displaystyle {\bf M}( p_1; p_2)\equiv T(s,\vec q) =\frac{is}{2m^{2}}\int d^{2}b\ e\,^{\displaystyle i\vec q\cdot \vec b}\ \bigl(1-e^{\displaystyle i{\mathbb X}(s,\vec b)}\,\bigr)
\end{equation}
where ${\mathbb X}(s,\vec b)$ is the eikonal function, written below, appropriate to the scattering when $s=(p_1+p_2)^2$, $\vec b$ is the impact parameter of the collision in the center of mass frame introduced in section III.B, $\vec q$ is the momentum transfer in that frame : $\vec q^{\,2} = -(p_1-p'_1)^2\ll s$, and $m$ the mass of the quark/antiquark.

Starting from (\ref{ELEQ}) and (\ref{EL}), the non trivial part of (\ref{Tsq}) reads, in the eikonal and quenched approximation : 
\begin{equation}\label{Xsq}
\displaystyle e^{\displaystyle i{\mathbb X}(s,\vec b)} = \mathcal{N}\int\frac{\mathrm{d}[\chi]}{\sqrt{\det(f\!\cdot\!\chi)}}\,e\,^{\displaystyle{ \frac{i}{4} \int{ \chi^{2} }} }\,e\,^{\displaystyle-{i\over2} \int\mathrm{d}^4z\ Q(z) \,(gf\!\cdot\!\chi)^{-1}(z)\,Q(z)}
\end{equation}

In the center of mass frame and in the large $g$ regime, where only the $R_{\mu}^a$ part of $Q_{\mu}^a$ is kept, this amplitude becomes,
\begin{equation}\label{Xsq1}
\displaystyle e^{\displaystyle i{\mathbb X}(s,\vec b)} = \mathcal{N}\int\frac{\mathrm{d}^8\chi}{\sqrt{\det(f\!\cdot\!\chi)}}\,e\,^{\displaystyle{ \frac{i}{4} \,\delta^4\chi^{2} }} \,e\,^{\displaystyle i g \,\varphi(b) \,\Omega\,(f\!\cdot\!\chi)^{-1}\,\Omega}
\end{equation}
thanks to the Dirac 4--delta contained  in each $R$. As a result, one can replace the functional $\mathrm{d}[\chi]$ integral by an ordinary $\mathrm{d}^8\chi$, replace the  integral $\displaystyle\int\mathrm{d}^4x\,\chi^2(x)$ by $\delta^4\chi^2$ where $\delta^4$ is an infinitesimal  4--volume, use eq.(\ref{local}) in the last exponential and finally replace the  $\delta^{(2)}(\vec{b})$  by $\varphi(b)$. 

Making the substitution $\delta^2\chi\rightarrow\chi$ so that the new $\chi$ is dimensionless, calling ${\cal R}$ the magnitude of $f\!\cdot\!\chi$ and averaging over the angular variables \cite{QCD1, QCD-II, QCD6}, one obtains for (\ref{Xsq1}) :
\begin{equation}\label{Xsq2}
\displaystyle e^{\displaystyle i{\mathbb X}(s,\vec b)} = N\int_0^{\infty}\frac{{\cal R}^7\mathrm{d}{\cal R}}{\sqrt{{\cal R}^8}}\,e\,^{\displaystyle{ \frac{i}{4}{\cal R}^{2} }} \,e\,^{\displaystyle i g \,\delta_q^2\,\varphi(b) \,{\cal R}^{-1}} = N\int_0^{\infty}\!\!{\cal R}^3\mathrm{d}{\cal R}\,e\,^{\displaystyle{ \frac{i}{4} {\cal R}^{2} }} \,e\,^{\displaystyle i g \,\delta_q^2\,\varphi(b) \,{\cal R}^{-1}}
\end{equation}
where the $\delta$  parameter is now called $\delta_q$ (\cite{QCD5'} sect.4) and the normalization constant $N$ is such that $\displaystyle e^{\displaystyle i{\mathbb X}(s,\vec b)} = 1$ when $g = 0$. 

The exponent $g \,\delta_q^2\,\varphi(b) \,{\cal R}^{-1}$ can be associated to the graph of  Fig.\ref{quark_gluonbundle}. There is a $g\,\delta_q$ attached at each quark--gluon bundle vertex, a $g^{-1}$ coming from the gluon bundle $(g{\cal R})^{-1}$, and a $\varphi(b)$ linked with this gluon bundle.

The binding potential $V(\vec r)$ between two quarks/antiquarks is related to the eikonal function by the formula (\cite{Eik}, chap.8) :
\begin{equation}\label{pot}
\displaystyle{\mathbb X}(s,\vec b) = - \int_{-\infty}^{+\infty}{\rm d}z\,V(\vec b + z\,\hat p_L)
\end{equation}
where $\hat p_L$ is a unit vector in the direction of longitudinal motion.

For small $\varphi(b)$,
\begin{equation}\label{lin}
\displaystyle i{\mathbb X}(b) = \ln(\varphi(b)) + \cdots
\end{equation}

The part of the eikonal function one is interested in is purely imaginary, as it should be, in order to obtain a binding potential, which itself is the imaginary part of the total potentiel~: ${\cal V} = V_S - iV_B$ where $V_S$ is the scattering potential. And here, $V = V_B$.

Using (\ref{deltaphy}), that is
$\varphi(b) = \varphi(0)\,e^{\displaystyle-(\mu b)^{2+\xi}}$, one obtains,
\begin{equation}\label{lin2}
\displaystyle i{\mathbb X}(b) = \displaystyle-(\mu b)^{2+\xi} + \cdots
\end{equation}
The binding potential can be recovered from (\ref{pot}) by taking the Fourier transform of ${\mathbb X}$. For small enough $\xi$, one gets (\cite{QCD-II}, sect.7) :
\begin{equation}\label{pot2}
V(r) \simeq \displaystyle \xi\,\mu\,(\mu r)^{1+\xi}
\end{equation}
which justifies the presence of this small $\xi$ parameter in the definition of $\varphi(b)$, so as to obtain a non zero potential. It can be shown (\cite{QCD-II} (72)--(77)) that this binding potential has the same form for baryons. For negative $\zeta$-values, interesting relations of (\ref{pot2}) to \emph{Lowest Landau Levels} \cite{Thess2}, to a non-commutative geometrical aspect of the transverse scattering plane, and to a \emph{Levy-flight} mode of propagation of confined quarks, seem to show up.

\subsection{Estimation of the light quark mass}
Using the confining potential $V(r)$ (\ref{pot2}), one can give an order of magnitude of the nucleon quark mass  by finding the ground state energy of a `` model pion '' first quantized hamiltonian (\cite{QCD-II} sect.8) relying on Heisenberg inequality and classical arguments. The result being of course~: ${\cal E}_0  = m_{\pi} $.

The hamiltonian of this $q$--$\bar q$ system at rest is :
\begin{equation}\label{pion}
\displaystyle{\cal H}_{\pi} = \displaystyle2m + {p^2\over m} + V(r)
\end{equation}
where $m$ is the mass of the quark/antiquark, and $r$ and $p$ the canonical pair defining the relative position and momentum of the system. One finds the minimal eigenvalue of ${\cal H}_{\pi}$ by using the Heisenberg inequality $p\ge 1/r$, followed by $\displaystyle {d{\cal H}_{\pi}\over dr}\bigg|_{r=r_0}\!\!\!= 0$. Replacing this $r_0$ value in the hamiltonian and assuming $\xi\ll1$, one has : 
\begin{equation}\label{Emin}
\displaystyle E_0 = \displaystyle2m + 3\mu\,\biggl({\xi^2\mu\over 4m}\biggr)^{1/3} 
\end{equation}

$E_0$ can be minimized again  when considered as a function of $x = \displaystyle{\mu\over m}$ :
\begin{equation}\label{Eminx}
\displaystyle E_0 = \displaystyle\mu\,\Bigl( \,{2\over x} + 3\,\Bigl({\xi\over2}\Bigr)^{2/3}\,x^{1/3}\,\Bigr) 
\end{equation}

The minimal value is obtained for $x = \displaystyle{2^{5/4}\over\sqrt\xi}\cdot$ and one then simply gets,
\begin{equation}\label{E0}
\displaystyle {\cal E}_0 \simeq 8m 
\end{equation}

It follows that $m \simeq \displaystyle{m_{\pi}\over8} \simeq 15\,$MeV/$c^2$ to be compared to the ``current'' mass \cite{PDG}, 

$m_u\simeq m_d\simeq 4-5\,$MeV/$c^2$.

One can now give a more precise meaning to the $EL$-$\mu$ parameter by assuming it is related to the pion mass, $\mu\simeq m_{\pi}$. This value, used in the following sections, leads to a small value of the deformation parameter, $\xi = 0.088$.

\subsection{Nucleon--nucleon binding potential}

\begin{figure}
\centering
\includegraphics[width=10cm]{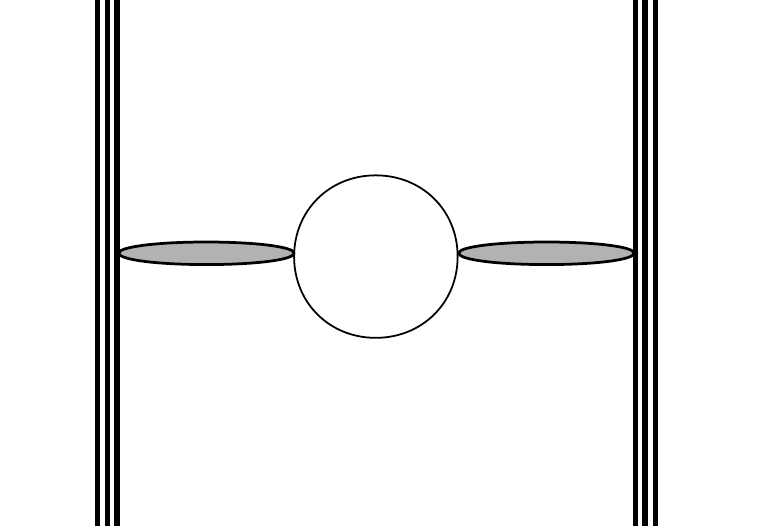}
\caption{A quark loop exchange through two gluon bundles ( shaded ovals ) between two quarks of two nucleons ( solid lines ).}
\label{oneloop}      
\end{figure}

One can't expect to find a nucleon--nucleon binding potential using the same eikonal as in the previous section. For, first, the distance between nucleons in a nucleus must be larger than that of quarks within a hadron, which means that $\varphi(b)$ must be replaced by a function with a larger $b$ range. And secondly, one would expect to find a link between this nuclear potential and the pion exchange mechanism inside nucleus, which a single gluon bundle exchange between quarks can't achieve. Taking the fermionic determinant $L[A]$ into account, (\ref{surprising}), will lead to a potential satisfying these two requirements.

Due to the relative complexity of the computations for arriving at useful formulas, the results are given without details, those details being supplied in \cite{QCD5}, \cite{QCD5'}, \cite{HerbSFG} chap.12,  and~\cite{ppscat}.

To begin with, the nucleon--nucleon binding process will be described as a two body interaction as can be shown in \cite{QCD5}, Eqs.(5) to (12).

Then, one assumes that the eikonal amplitude  leading to the nuclear potential relies on the $L[A]$ contribution only. Writing (\ref{surprising}) in the eikonal approximation and neglecting the action of the linkage operator on the fermionic determinant, one obtains :
\begin{equation}\label{XL}
\displaystyle e^{\displaystyle i{\mathbb X}(s,\vec b)} = \mathcal{N}\int\frac{\mathrm{d}[\chi]}{\sqrt{\det(f\!\cdot\!\chi)}}\,e\,^{\displaystyle{ \frac{i}{4} \int{ \chi^{2} }} }\,e\,^{\displaystyle{L}[(gf\!\cdot\!\chi)^{-1}( R_1 + R_2 )] }
\end{equation}
where $R_{1,2}$ are the quark--currents (\ref{QC}) involved in the binding and where $L[A]$ is given by (\ref{Eq:LFradkin01}), its ordered exponential being coped with the help of (\ref{full}).
Neglecting spin effects, it follows that the `` useful '' part of $L[A]$ is :
\begin{equation}\label{LU}
\displaystyle  e\,^{\displaystyle - i g \int_0^{s}ds' \, u'^{\mu}(s) \, \Omega_i^{a}(s) \, A^{a}_{\mu}(y_i-u_i(s)) } - 1
\end{equation}

Keeping only the first term of the exponential ( one fermion loop ) in the eikonal approximation, and integrating over the angular variables, one finds (\cite{QCD5}, Eqs.(40), and \cite{ppscat} Eq.(19)),
\begin{equation}\label{Xsq3}
\displaystyle e^{\displaystyle i{\mathbb X}(s,\vec b)} = N\int_0^{\infty}\!\!{\cal R}_1^3\,\mathrm{d}{\cal R}_1\,e\,^{\displaystyle{ \frac{i}{4} {\cal R}_1^{2} }}\,\int_0^{\infty}\!\!{\cal R}_2^3\,\mathrm{d}{\cal R}_2\,e\,^{\displaystyle{ \frac{i}{4} {\cal R}_2^{2} }} \,e\,^{\displaystyle i \,{\cal C}(s,\vec b)\,{\cal R}_1^{-1}{\cal R}_2^{-1}}
\end{equation}
with :
\begin{equation}\label{Csb}
\displaystyle  {\cal C}(s,\vec b) = g^2\,\delta_q^2\,\bigl({\kappa\over \bar\mu^2}\bigr)\,\Delta\bar\varphi(b)
\end{equation}

A few words of explanation are appropriate : 

The ${\cal C}\,{\cal R}_1^{-1}{\cal R}_2^{-1}$ factor can be diagrammatically depicted by  Fig.\ref{oneloop}.  There is a $g\,\delta_q$ attached to each `` physical '' quark--gluon bundle vertex, a $g\,\delta$ attached to each end of the `` loop  '' quark--gluon bundle vertex, this $\delta$ tending to 0, together with a logarithmically divergent quantity $l$ associated to the quark loop diagram, such that $\displaystyle\delta^2\,l = {\kappa\over \bar\mu^2}$ is finite (\cite{QCD5'}, Sect.4) and each of the two gluon bundles gives a $(g\,{\cal R}_i)^{-1}$ where ${\cal R}_i$ is the magnitude of $f\!\cdot\!\chi_i$. Concerning the transverse fluctuation function $\bar\varphi(b)$, it is obtained as the convolution product of the two $\varphi(b)$, one for  each gluon bundle, the result being a function with a larger dispersion (\cite{QCD5}, Eq.(26)). The $\bar\mu$ mass needs not being the same as in $\varphi$. The last ingredient in (\ref{Csb}) is the laplacian in front of $\bar\varphi$, induced by the quark loop (\cite{QCD5}, Eqs.(30) to (40)). 

As a last comment, Fig.\ref{oneloop} can be seen as the exchange of a virtual $\pi$ between the nucleons, via two gluon bundles.

From this, one can compute a ( very qualitative ) binding potential following a similar path as in Subsection {\bf A} above. The difference being that one expands to first order the left hand side of eq.(\ref{Xsq2}) and in its right hand side the exponential factor containing ${\cal C}(s,\vec b)$, as ${\mathbb X}$ and ${\cal C}$ are both small quantities in this binding process; moreover, and the small $\xi$ parameter of $\varphi$ is no longer essential and can be neglected. One finds (\cite{QCD5}, Eq.(48)),
\begin{equation}\label{pot3}
V(r) \simeq \displaystyle g^2\,\bar\mu\,( 2 - \bar\mu^2\,r^2 )\,e\,^{\displaystyle -{\bar\mu^2r^2\over2}}
\end{equation}

This potential can not be expected to be reliable at small $\bar\mu\,r$ values, where the approximation of one quark loop exchange is no more valid.

\subsection{Estimation of the size of the deuteron}

Using the potential (\ref{pot3}), a deuteron model can be built, starting from the classical two equal--mass nucleon hamiltonian :
\begin{equation}\label{deuteron}
\displaystyle{\cal H}_D =  {p^2\over m_N} + V_0\,( 1 - {\bar\mu^2\,r^2\over2} )\,e\,^{\displaystyle -{\bar\mu^2r^2\over2}}
\end{equation}

With the help of the Heisenberg inequality $\displaystyle p\ge {1\over r}$, one can infer the size of the deuteron in its fundamental state.

Defining the dimensionless variable $x = \bar\mu r$, the hamiltonian can be rewritten :
\begin{equation}\label{deuteron1}
\displaystyle{\cal H}_D =  {\bar\mu^2\over m_N\,x^2} + V_0\,( 1 - {x^2\over2} )\,e\,^{\displaystyle -{x^2\over2}}
\end{equation}

If $\bar\mu$ can be taken as the pion mass $\displaystyle m_\pi\simeq 140$\,MeV/$c^2$. With $m_N \simeq 1$\,GeV/$c^2$, that leads to $\displaystyle M = {\bar\mu^2\over m_N}\simeq20$\,MeV/$c^2$.

The minimum of ${\cal H}_D$ is found by solving : $\displaystyle {d{\cal H}_D\over dx}\bigg|_{x=x_0}\!\!\!= 0$

That leads to the equation :
\begin{equation}\label{mindeut}
\displaystyle0 =  -{2M\over x_0^3} + {V_0\over2}\,( x_0^3 - 4x_0 )\,e\,^{\displaystyle -{x_0^2\over2}}
\end{equation}

Taking out the exponential and bringing it back in ${\cal H}_D$, one finds :
\begin{equation}\label{minE}
\displaystyle E_0 =  {x_0^4 - 6x_0^2 + 4\over x_0^4\,( x_0^2 - 4  )}\,M 
\end{equation}

Solving this equation for $x_0$ using $E_0\simeq -2$\,MeV, one finds $x_0\simeq 0.85$ that leads to an unphysical negative $V_0$ and $x_0\simeq 2.2$ that leads to a size for the deuteron of $r_0\simeq 3$ fm, which is a correct order of magnitude. 

\subsection{Application to $pp$ elastic scattering}

\begin{figure}
\centering
\includegraphics[width=10cm]{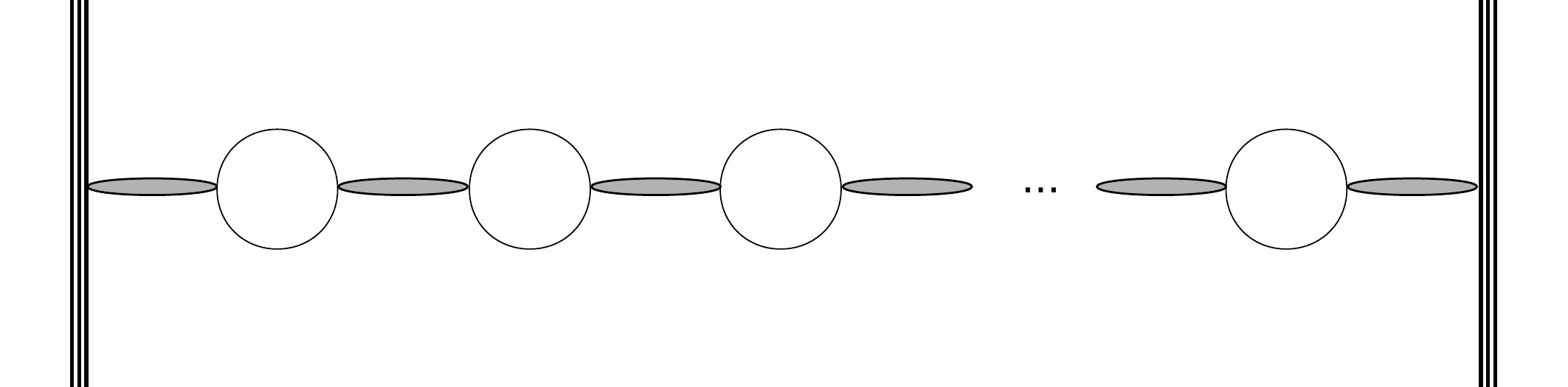}
\caption{A linear chain of gluon bundles and quark loops exchange between two quarks of two nucleons. No more than two bundles can  be  attached to a quark loop \cite{ppscat}.}
\label{chainloops}      
\end{figure}

The two physical ingredients of our eikonal non--perturbative QCD theory, the exchange of gluon bundles, schematized in Fig.\ref{quark_gluonbundle} for one bundle exchange and the exchange of quark loop chains schematized in Fig.\ref{oneloop} for one quark loop chain exchange, that have been used in binding contexts, can also be employed for the description of proton--proton elastic scattering, where the eikonal approximation is justified.

Before putting forward a theoretical cross--section for these processes, one has to recall a major result of these Effective Locality induced dynamics : only  a few amplitudes -- in the eikonal context -- involving quarks, gluon bundles and linear quark loop chains (see Fig.\ref{chainloops}) are non--zero. Only exchanges between two different quarks are allowed. They take the shape  of ladders if several of these `` building blocks '' are involved  (see Fig.\ref{gluonbundles}). In particular, no exchanges of bundles and chains can occur on a single quark line : there are no self--energy graphs for quarks in this non--perturbative QCD regime, all of these features established in Ref.\cite{QCD5'}.

\begin{figure}
\centering
\includegraphics[width=10cm]{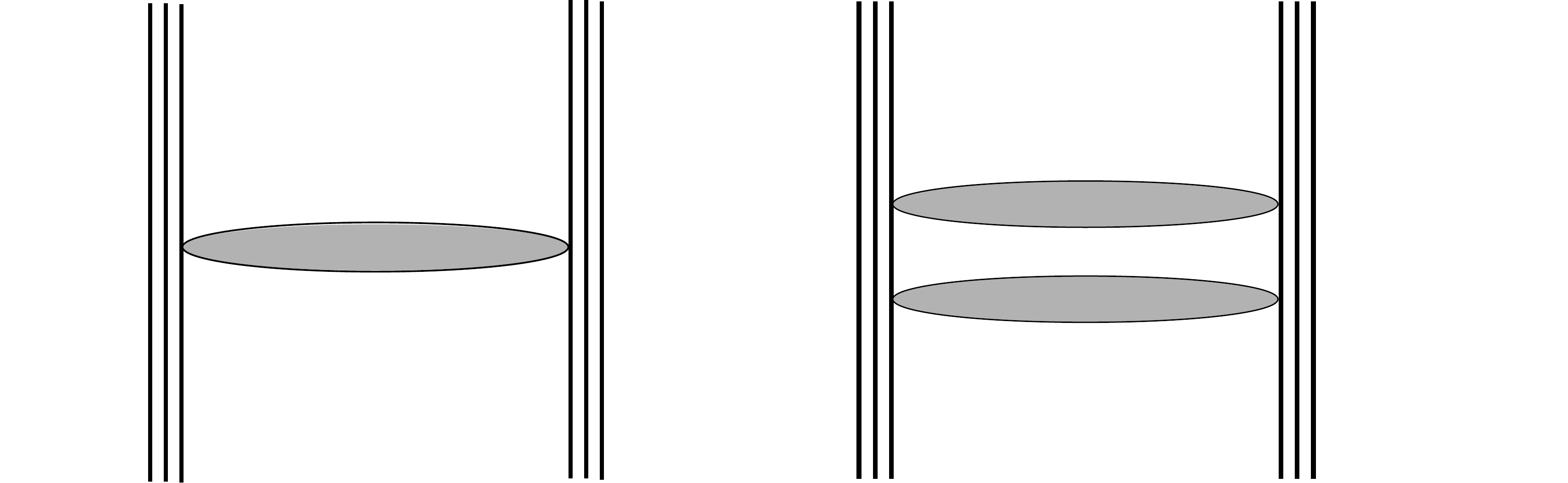}
\caption{One and two gluon bundles exchange between two quarks of two nucleons.}
\label{gluonbundles}      
\end{figure}

Turning to the elastic scattering between nucleons, the aim is to produce $QCD$ curves to compare to the old ISR \cite{isr_data1,isr_data2,isr_data3,isr_data4} and the recent LHC TOTEM \cite{LHC} $pp$ elastic scattering data. 

The appropriate starting point is again the eikonal amplitude (\ref{Tsq}) :
\begin{equation}
\displaystyle {\bf M}( p_1; p_2)\equiv T^{pp}(s,\vec q) =\frac{is}{2M^{2}}\int d^{2}b\ e\,^{\displaystyle i\vec q\cdot \vec b}\ \bigl(1-e^{\displaystyle i{\mathbb X}^{pp}(s,\vec b)}\,\bigr)
\end{equation}
with $s = 4\,{\cal E}^2$ where ${\cal E}$ is the center of mass energy of each proton, $M$ the proton mass and $\vec q$ the momentum transfer in that frame : $\vec q^2 = |t|$.

A description of the elastic data involves exchanges of both gluon bundles and quark loop chains. Obviously, this description will be crude, due to all the approximations made along the computations, the neglect of spin effects being one of them. Still one will get curves whose shapes qualitively reproduce the data features. Needless to recall that this approach is strictly based on the $QCD$ lagangian (\ref{1}), from which \emph{Pomerons} are absent. 

Recalling the gluon bundle exchange amplitude (\ref{Xsq2}),
\begin{equation}\label{Xsq4}
\displaystyle e^{\displaystyle i{\mathbb X}^{(GB)}(s,\vec b)} = N\int_0^{\infty}\!\!{\cal R}^3\mathrm{d}{\cal R}\,e\,^{\displaystyle{ \frac{i}{4} {\cal R}^{2} }} \,e\,^{\displaystyle i g \,\delta_q^2\,\varphi(b) \,{\cal R}^{-1}}
\end{equation}
and the quark loop chain exchange amplitude, Eqs.(\ref{Xsq3}) and (\ref{Csb}),
\begin{equation}\label{Xsq5}
\displaystyle e^{\displaystyle i{\mathbb X}^{(QLC)}(s,\vec b)} = N\int_0^{\infty}\!\!{\cal R}_1^3\,\mathrm{d}{\cal R}_1\,e\,^{\displaystyle{ \frac{i}{4} {\cal R}_1^{2} }}\,\int_0^{\infty}\!\!{\cal R}_2^3\,\mathrm{d}{\cal R}_2\,e\,^{\displaystyle{ \frac{i}{4} {\cal R}_2^{2} }} \,e\,^{\displaystyle i \,g^2\,\delta_q^2\,\bigl({\kappa\over \bar\mu^2}\bigr)\,\Delta\bar\varphi(b)\,{\cal R}_1^{-1}{\cal R}_2^{-1}}
\end{equation}
and integrating over the $\cal R$s (\cite{ppscat}, Eqs.(20)--(23)), one obtains :
\begin{equation}\label{XGB}
\displaystyle e^{\displaystyle i{\mathbb X}^{(GB)}(s,\vec b)} =  \,e\,^{\displaystyle {{\sqrt i}\over2}\, g \,\delta_q^2\,\varphi(b)}
\end{equation}
and
\begin{equation}\label{XQLC}
\displaystyle e^{\displaystyle i{\mathbb X}^{(QLC)}(s,\vec b)} = \,e\,^{\displaystyle {1\over4}\,g^2\,\delta_q^2\,\bigl({\kappa\over \bar\mu^2}\bigr)\,\Delta\bar\varphi(b)}
\end{equation}

That leads to,
\begin{eqnarray}\label{Tpp}
\displaystyle T^{pp}(s,\vec q) &=&\frac{is}{2M^{2}}\int d^{2}b\ e\,^{\displaystyle i\vec q\cdot \vec b}\ \bigl(1-e^{\displaystyle i{\mathbb X}^{(GB)}(s,\vec b)}\,e^{\displaystyle i{\mathbb X}^{(QLC)}(s,\vec b)}\,\bigr)\nonumber\\  &=&\frac{is}{2M^{2}}\int d^{2}b\ e\,^{\displaystyle i\vec q\cdot \vec b}\ \bigl(1-e\,^{\displaystyle {{\sqrt i}\over2}\, g \,\delta_q^2\,\varphi(b)}\,e\,^{\displaystyle {1\over4}\,g^2\,\delta_q^2\,\bigl({\kappa\over \bar\mu^2}\bigr)\,\Delta\bar\varphi(b)}\,\bigr)
\end{eqnarray}

The elastic differential cross--section is given by (\cite{Eik}, Chaps. 8 and 10),
\begin{equation}\label{sig}
\displaystyle\frac{d\sigma^{pp}}{dt} = \frac{M^4}{\pi s^2}\, | T^{pp} |^2
\end{equation}

Before computing $| T^{pp} |^2$, one has to specify first the transverse fluctuation functions :
\begin{equation}\label{fifi}
\varphi (b) = {m^2\over\pi}\, e^{\displaystyle -m^2b^2},\ \ \ \ \bar\varphi (b) = {\bar m^2\over2\pi}\, e^{\displaystyle -{\bar m^2\over2}b^2}\end{equation}
where $\varphi$ and $\bar\varphi$ are normalized, second the values of $m$ and $\bar m$ that depend on the energy of the reaction but should be close to the pion mass, and finally give a physical definition of $\delta_q$ that was not necessary in the previous sub--sections but is now, in this scattering frame where the energy of the reaction is a fundamental parameter : 
\begin{equation}\label{delq}
\displaystyle\delta_q  = \bigl({\lambda\over m}\bigr)\, \bigl({m\over E}\bigr)^p
\end{equation}
with $E$ the energy of each quark involved in the scattering, $\displaystyle E ={\cal E}/3$, $p$ is a small positive parameter accounting for the decrease of the cross--section with increasing energy, and $\lambda$ dimensionless and small. All these quantities depend smoothly on $E$.

If one considers an amplitude where one gluon bundle and one quark loop chain is exchanged between quarks, then one has to expand each exponential to first order only and from (\ref{XGB}), (\ref{XQLC}) and (\ref{Tpp}) one obtains,
\begin{equation}\label{Tpp1}
\displaystyle T_1^{pp}(s,\vec q) = -\frac{is}{2M^{2}}\int d^{2}b\ e\,^{\displaystyle i\vec q\cdot \vec b}\ \Bigl(\,\displaystyle {{\sqrt i}\over2}\, g \,\delta_q^2\,\varphi(b) + \displaystyle {1\over4}\,g^2\,\delta_q^2\,\bigl({\kappa\over \bar\mu^2}\bigr)\,\Delta\bar\varphi(b)\Bigr)
\end{equation}

Adding to this $T_1$ amplitude a two gluon bundle exchange (see Fig.\ref{gluonbundles}) by expanding ${\mathbb X}^{(GB)}$ to second order, one gets : 
\begin{equation}\label{Tpp2}
\displaystyle T_1^{pp}(s,\vec q) = -\frac{is}{2M^{2}}\int d^{2}b\ e\,^{\displaystyle i\vec q\cdot \vec b}\ \Bigl(\,\displaystyle {{\sqrt i}\over2}\, g \,\delta_q^2\,\varphi(b) + \,{1\over2}\,{i\over4}\,g^2 \delta_q^4\,\varphi^2(b) + \displaystyle {1\over4}\,g^2\,\delta_q^2\,\bigl({\kappa\over \bar\mu^2}\bigr)\,\Delta\bar\varphi(b)\Bigr)
\end{equation}

That leads to two cross--sections, $\displaystyle\frac{d\sigma_1^{pp}}{dt}$ and $\displaystyle\frac{d\sigma_2^{pp}}{dt}$ whose expressions are not too illuminating, and are given in \cite{ppscat} eq.(35) and (36).

\begin{figure}
\centering
\includegraphics[width=10cm]{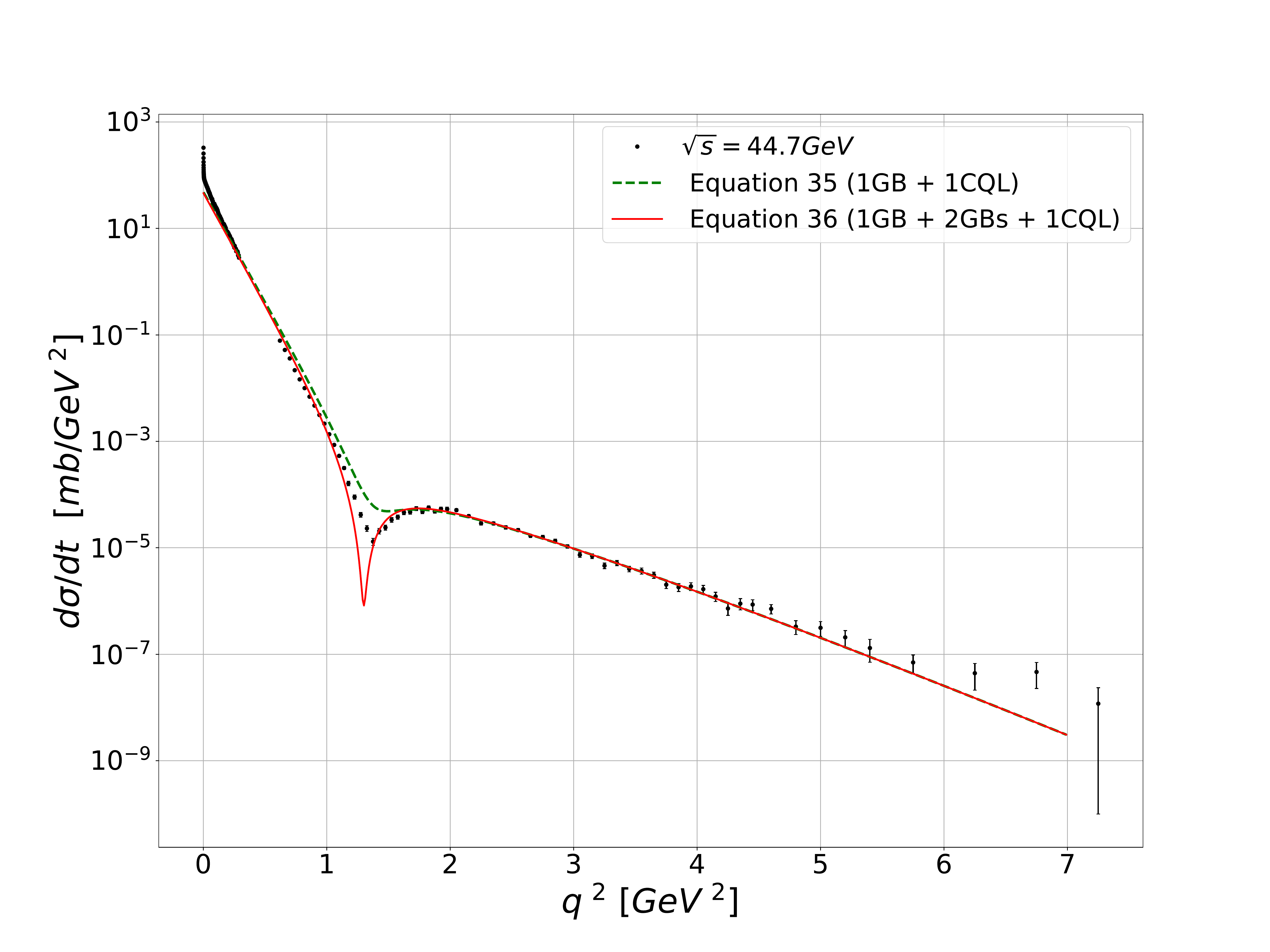}
\caption{Elastic $pp$ scattering differential cross--section at $\sqrt s = 44.7$ GeV. Black dots are experimental data, dashed line is the result of \cite{ppscat}, Eq.(35), solid line comes from \cite{ppscat} eq.(36).}
\label{44_7gev}      
\end{figure}

\begin{figure}
\centering
\includegraphics[width=10cm]{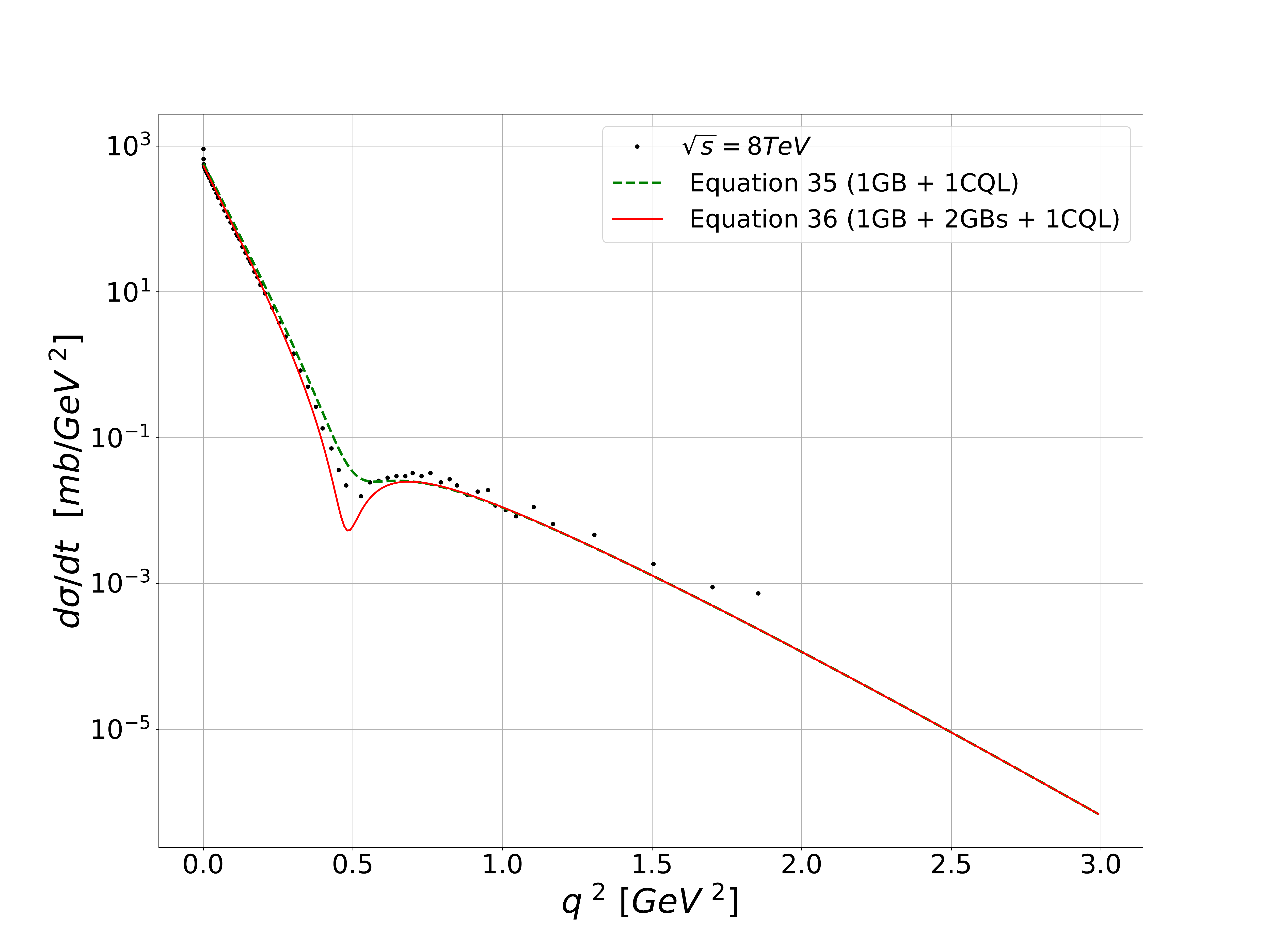}
\caption{Elastic $pp$ scattering differential cross--section at $\sqrt s = 8$ TeV. Black dots are experimental data, dashed line is the result of \cite{ppscat} eq.(35), solid line comes from \cite{ppscat}, Eq.(36).}
\label{8tev}      
\end{figure}

Two examples of the curves obtained using (\ref{Tpp1}) and (\ref{Tpp2}) are given in Fig.\ref{44_7gev} for the ISR data and Fig.\ref{8tev} for TOTEM data.

A necessary comment on those QCD results is that the gluon bundle contribution to the cross--section ( one or two exchanges ) comes out as an $e\,^{\displaystyle -{q^2\over4m^2}}$ that gives the low $q^2$ part of the curves, below the `` dip " shown by the data, while the one quark loop chain, thanks to the laplacian in front of $\bar\varphi(b)$ leads to a term proportionnal to $q^2\,e\,^{\displaystyle -{q^2\over4m^2}}$ that gives the major contribution  for the cross--section for larger $q^2$, above the `` dip ", the sum of the two ( bundles and loop ) qualitatively accounting for this experimental dip. It is remarkable that just the two or three first terms of the eikonal exponential are sufficient to give a qualitatively fair description of data covering a range of energy between 30  GeV to 10 TeV.

\section {\label{SEC:5}Conclusion}
Effective Locality has been derived some years ago in the context of Lagrangian quantum field theory for the sake of exploring the non-perturbative sector of $QCD$. Its fundamental step consists in `integrating out' the gluonic degrees of freedom of the original generating functional of $QCD$. Quantization is carried out by relying on functional differentiations rather than functional integrations, the two procedures of quantization being equivalent whenever the \emph{Wick theorem} applies to time-ordered products of quantum field operators.
\par
By adding and subtracting Lagrangian density terms ordinarily used as gauge-fixing conditions in the functional integration context, invertible gauge field- functions ({\textit{i.e.}} propagators) of any sort are generated which render perfectly well-defined all of the standard operations of functional differentiation which complete the quantization procedure. 
\par
In the end, it is really remarkable that gluonic degrees disappear to the exclusive benefit of rank~2 covariant tensor fields endowed with color indices, the $\chi^a_{\mu\nu}$-terms, which are introduced in order to `linearize' the original $F^2$-non abelian field-strength tensor of the $QCD$ Lagrangian density. 
\par
Somewhat similar in form, an astounding result was derived 30 years ago in the pure Yang Mills theory through an \emph{instanton} calculation, that is, in the euclidean case. At leading order of a semi-classical expansion, the $\chi^a_{\mu\nu}$-fields could consistently be understood as the original field strength tensors evaluated on instanton $A^a_\mu$- gauge field configurations, $\chi^a_{\mu\nu}=F^a_{\mu\nu}(A^b_{inst.})$. Now, besides the euclidean framework of this derivation, functional integration was used to quantize the Yang Mills generating functional, with, as a consequence gauge-fixing conditions transferred from the original $A^a_\mu$- gauge fields to the new $\chi^a_{\mu\nu}$-field variables, with ultimately, a \emph{Gribov's copies issue} met again. 
\par
On the contrary, effective locality calculations take place in Minkowski spacetime right from the onset, avoid any gauge-fixing procedure and related Gribov's copy problem, and by construction, preserve the full original non-abelian gauge invariance of $QCD$. This certainly stands for the most striking and physically interesting aspect of Effective Locality.
\par\medskip
In this article, a summarized review of eight, among the theoretical aspects and consequences of the Effective Locality property have been presented out of an ongoing list, and likewise, five phenomenological applications are also given. If the latter are often obtained on the bases of rough approximations to the exact expressions, it remains that most of them can be motivated and that they are found in line with the expected features of the non-perturbative regime of $QCD$. If, as Wisdom claims, it is true that \emph{When an idea is a good one, it is fecund at tree-level}, then these first results could be considered as encouraging.
\par\bigskip\bigskip\noindent
{\textbf{Conflicts of Interest:}} The authors declare no conflicts of interest.


\begin{thebibliography}{**} 
\bibitem{QCD1}
 H.M. Fried, Y. Gabellini, T. Grandou and  Y.-M. Sheu,  Eur. Phys. J. C \textbf{65}, 395 (2010).
\bibitem{QCD-II}
H.M. Fried, T. Grandou and Y.-M. Sheu,  Ann. Phys. {\bf{327}}, 2666 (2012).

\bibitem{QCD5}
H.M. Fried, Y. Gabellini, T. Grandou and Y.-M. Sheu, 
Ann. Phys.{\textbf{ 338}}, 107 ( 2013). 
 
\bibitem{QCD6} H.M. Fried, T. Grandou and Y.-M. Sheu, Ann. Phys. {\bf{344}}, 78 (2014).

\bibitem{QCD5'}  H.M. Fried, P.H. Tsang, Y. Gabellini,  T. Grandou and Y.-M. Sheu, Ann. Phys.{\textbf{ 359}}, 1 (2015).

\bibitem{RefF}
H. Reinhardt, K. Langfeld and L. von Smekal, Phys. Lett. B {\bf{300}}, 111 (1993); H. Reinhardt,``Dual description of QCD"; arXiv:hep-th/9608191v1 (1996).

\bibitem{SW1994} N. Seiberg and E. Witten, Nucl. Phys. B {\textbf{426}}, 19 (1994); arXiv:hep-th/9407087
24 [Erratum-ibid. B {\textbf{430}}, 485 (1994)].

\bibitem{Herb} H.M. Fried, \textit{Green's Functions and Ordered Exponentials}, Cambridge University Presss (2002).

\bibitem{IZ} See, for instance, C. Itzykson and J.B. Zuber, \textit{Quantum Field Theory}, McGraw-Hill Inc. (1980) chap. 4

\bibitem{Halpern1977a}
M.B. Halpern, Phys. Rev. D \textbf{16}, 1798  (1977).

\bibitem{Halpern1977b}
M.B. Halpern, Phys. Rev. D \textbf{16}, 3515  (1977).





\bibitem{RefO} A. Zee, \textit{Quantum Field Theory in a Nutshell, 2nd Edition}, Princeton University Press (2010) chap.I.3.

\bibitem{RefI}
 T. Grandou, EPL {\textbf{107}}, 11001 (2014). H.M. Fried, T. Grandou and R. Hofmann, Int. J. of Mod. Phys. A
\textbf{ 31}, Nos. 20 \& 21, 1650120 (2016); arXiv:1504.05502 [hep-th].

\bibitem{RT} R F. Streater  and A S. Wightman  \textit{PCT, Spin and Statistics, and All That}, Princeton University Press (2000). ISBN: 0691070628

\bibitem{prepara} Y. Gabellini and T. Grandou, work in completion.

\bibitem{Eik} H.M. Fried, \textit{Basics of Functional Methods and Eikonal Models}, Editions Fronti\`eres (1990). 



\bibitem{tg} H.M. Fried, T. Grandou and R. Hofmann, Mod. Phys. Lett. A {\bf32}, 1730030 (2017);
arXiv:1706.02264v1 [hep-th].

\bibitem{Lavelle1996}
M. Lavelle, \textit{Construction and Consequences of Coloured Charges in QCD}, Proceedings of Workshop on \textit{Quantum Chromodynamics: Collisions, Confinement and Chaos}, B. Muller and H.M. Fried (Eds), American University of Paris, 3-8 June 1996; arXiv:hep-ph/9706521.


\bibitem{Stan} S.J. Brodsky, J.R. Ellis and M. Karliner, Phys. Lett. B {\textbf{206}} (1988) 309; G.F. de T\'eramond and S.J. Brodsky, e-Print:  2103.10950 [hep-ph].


\bibitem{Matveev2002}
V.A. Matveev, V.I. Savrin, A.N. Sissakian and A.N. Tavkhelidze, Theor. Math. Phys. {\bf{132}}, 1119 (2002).

\bibitem{Ted}
G.W. Johnson and M.L. Lapidus, \textit{The Feynman Integral and Feynman's Operational Calculus}, Oxford University Press (2000).

\bibitem{GR} I.S. Gradshteyn and I.M. Ryzhik, \textit{Tables of Integrals, Series and Products, 4th Edition}, Academic Press, London (1980).

\bibitem{pomme}
A. Apelblat, \textit{Table of Definite and Infinite Integrals}, Elsevier Science Ltd. (1983), p.26, formula 68.

\bibitem{Ferrante}
  D.D. Ferrante, G.S. Guralnik,  Z. Guralnik and C. Pehlevan, {BROWN-HET-1611} (2011).
  
  \bibitem{Dmitrasinovic} V. Dmitrasinovic, Phys. Lett. B {\bf{499}}, 135 (2001).
  
  \bibitem{Nieuwenhuizen}  G.C. Nayak and P. van Nieuwenhuizen, Phys. Rev. D {\bf{71}}, 125001 (2005).
  
  \bibitem{Nayak} G.C. Nayak, Phys. Rev. D {\bf{72}}, 125010 (2005).
  
  \bibitem{Rhodos} T.G. and R. Hofmann, Mod. Phys. Lett. {\bf{A}}35, (2020),  2050230; arXiv:2001.08136v1[hep-th].
  
  \bibitem{Kleinert} H. Kleinert, \textit{Particles and Quantum Fields}, World Scientific (2016) p.602.
  
  \bibitem{prepar} Y. Gabellini and T. Grandou, in preparation.
  
  
\bibitem{Trento}
G. van Baalen, D. Kreimer, D. Uminsky and K. Yeats, Ann. Phys. {\bf{325}}, 300 (2010).

\bibitem{Thess} A. Maas, \textit{XIIth Quark Confinement and the Hadron Spectrum}, Thessaloniki, Greece (2016).

\bibitem{Dyakonov} D.I. Dyakonov and V.Yu. Petrov, Phys. Lett. B {\textbf{147}},  351 (1984).

\bibitem{tgpt} T. Grandou and P.H. Tsang, Mod. Phys. Lett. {\bf{A}}, 

https://doi.org/10.1142/S0217732319503358; http://arxiv.org/abs/1905.05666; Physica~Scripta.

\bibitem{FGHF}F. Guerin and H.M. Fried, Phys. Rev. D {\bf{33}}, 3039 (1986).

\bibitem{Mehta}M.L. Mehta, \textit{Random Matrices}, Academic Press (1967).

\bibitem{Thess2} Cf. \textit{XIIth Quark Confinement and the Hadron Spectrum}, Thessaloniki, Greece (2016).




 






\bibitem{PDG} M. Tanabashi et \textit{al.} (Particle Data Group), Phys. Rev. D {\textbf{98}}, 030001 (2018) and 2019 update.

\bibitem{HerbSFG} H.M. Fried, \textit{Modern Functional Quantum Field Theory - Summing Feynman Graphs}, World Scientific (2014).

\bibitem{ppscat} H.M. Fried, P.H. Tsang, Y. Gabellini, T. Grandou and Y.M. Sheu, Int. J. Mod. Phys. A
{\bf 34}, 1950236 (2019).

\bibitem{isr_data2} C.N.P.S.B. Collab. (M. Ambrosio et al.), Phys. Lett. B 115, 495 (1982), \url{http://dx.doi.org/10.1016/0370-2693(82)90400-2}

\bibitem{isr_data3} A.B.C.D.H.W. Collab. (A. Breakstone et al.), Nucl. Phys. B 248, 253 (1984), \url{http://dx.doi.org/10.1016/0550-3213(84)90595-9}

\bibitem{isr_data4} N. Amos et al., Nucl. Phys. B 262, 689 (1985), \url{https://www.sciencedirect.com/science/article/pii/0550321385905115}.

\bibitem{isr_data1} U. Amaldi, K. Schubert, Nucl. Phys. B 166, 301 (1980), \url{http://dx.doi.org/10.1016/0550-3213(80)90229-1}.

\bibitem{LHC} TOTEM Collaboration:

Europhys. Lett. 95 (2011) 41001,  \url{https://doi.org/10.1209/0295-5075/95/41001}

Europhys. Lett. 101 (2013) 21002,  \url{https://doi.org/10.1209/0295-5075/101/21002} 

Europhys. Lett. 101 (2013) 21004 , \url{https://doi.org/10.1209/0295-5075/101/21004} 

Nucl. Phys. B 245, 275 (2013) \url{https://doi.org/10.1016/j.nuclphysbps.2013.10.054}

Eur. Phys. J. C 74, 3175 (2014) \url{https://doi.org/10.1140/epjc/s10052-014-3175-x} 

Eur. Phys. J. C 76, 661 (2016), \url{http://dx.doi.org/10.1140/epjc/s10052-016-4399-8}.






















\end{thebibliography}
\end{document}